\documentclass[conference]{IEEEtran}
\IEEEoverridecommandlockouts
\usepackage{cite}
\usepackage{amsmath,amssymb,amsfonts}

\usepackage{graphicx}
\usepackage{mathtools}

\usepackage[none]{hyphenat}
\usepackage{textcomp}
\usepackage{array}
\usepackage{kantlipsum}
\usepackage{flexisym}
\usepackage{stackengine}
\usepackage{booktabs}
\usepackage{lipsum}
\usepackage{multirow}
\usepackage[table,xcdraw]{xcolor}
\usepackage{xcolor}
\usepackage{caption}
\usepackage{subcaption}

\usepackage[mathscr]{euscript}

\DeclareSymbolFont{rsfs}{U}{rsfs}{m}{n}
\DeclareSymbolFontAlphabet{\mathscrsfs}{rsfs}
\usepackage[english]{babel}
\usepackage[utf8]{inputenc}

\usepackage{mathrsfs}
\usepackage[noend]{algpseudocode}
\usepackage{algpseudocode}

\DeclareFontFamily{U}{calligra}{}
\DeclareFontShape{U}{calligra}{m}{n}{<->callig15}{}

\usepackage{breqn}

\def\BibTeX{{\rm B\kern-.05em{\sc i\kern-.025em b}\kern-.08em
    T\kern-.1667em\lower.7ex\hbox{E}\kern-.125emX}}
\usepackage[bookmarks=false]{hyperref}

\begin{document}

\title{Physics-Informed Deep Learning and Partial Transfer Learning for Bearing Fault Diagnosis in the Presence of Highly Missing Data}

\author{Mohammadreza Kavianpour$^{1,*}$, Parisa Kavianpour$^2$, Amin Ramezani$^1$\\
$^1$Faculty of Electrical and Computer Engineering, Tarbiat Modares University, Tehran, Iran\\
$^2$Faculty of Technology and Engineering, University of Mazandaran, Babolsar, Iran\\
E-mail address: kavianpour@modares.ac.ir

}

\maketitle

\begin{abstract}
One of the most significant obstacles in bearing fault diagnosis is a lack of labeled data for various fault types. Also, sensor-acquired data frequently lack labels and have a large amount of missing data. This paper tackles these issues by presenting the PTPAI method, which uses physics-informed deep learning- based technique to generate synthetic labeled data. Labeled synthetic data makes up the source domain, whereas unlabeled data with missing data is present in the target domain. Consequently, imbalanced class problems and partial-set fault diagnosis hurdles emerge. To address these challenges, the RF-Mixup approach is used to handle imbalanced classes. As domain adaptation strategies, the MK-MMSD and CDAN are employed to mitigate the disparity in distribution between synthetic and actual data. Furthermore, partial-set challenge is tackled by applying weighting methods at the class and instance levels. Experimental outcomes on the CWRU and JNU datasets indicate that the proposed approach effectively addresses these problems.
\end{abstract}

\section{Introduction} \label{sec1}
Bearings are continually subjected to numerous faults as one of the primary components of rotating machinery owing to operating in a harsh environment and lengthy working duration. As a result, diagnosing bearing faults in real-world scenarios is critical and data-driven methods have become a prevalent solution for this purpose \cite{HE2023579, WU2023439, LI2023523}. Despite the widespread application of data-driven approaches in fault diagnosis, these methods typically require a large quantity of data for effective training. Nevertheless, the acquisition of substantial quantities of faulty data poses several challenges. Firstly, certain faults can result in the complete failure of the rotating machinery, making it impractical to restart the machinery and collect data under such conditions. Secondly, the deterioration process from healthy to breakdown is often lengthy. In this scenario, collecting faulty data is expensive. Thirdly, some faults may not have occurred until the data was collected, leading to a lack of training data for specific fault types. Lastly, most of the collected data consists of instances of the machinery in a healthy state, with a limited number of instances representing faulty conditions. \cite{rombach2023controlled, 9737184}.

\par Researchers have used transfer learning techniques to tackle this challenge in recent years \cite{kavianpour2021intelligent,GUO2024122806,LEE2024124084}. These techniques include subdomain adaptation (SDA) methods like local maximum mean discrepancy (LMMD) \cite{ghorvei2021unsupervised} and domain adaptation (DA) methods such as maximum mean discrepancy (MMD) \cite{li2024fault}, central moment matching (CMD) \cite{meng2022research}, correlation alignment (CORAL) \cite{9737160}, conditional adversarial domain adaptation (CDAN) \cite{misbah2023fault}, and adversarial domain discriminator \cite{chen2023collaborative}. In \cite{10251665}, a model with a 1-dimensional convolutional neural network (CNN) as its backbone is utilized, alongside the integration of the MMD criterion as a domain adaptation approach. In \cite{ghorvei2023spatial}, a hybrid transfer learning method is introduced that combines LMMD and adversarial learning to address the distribution discrepancy arising from changing working conditions. The work presented in \cite{zhou2023deep} proposes a semi-supervised adversarial domain adaptation method to alleviate the distribution disparity between synthetic and real data. In \cite{qian2023deep}, the authors utilize a technique that combines MMD and CORAL to mitigate the distribution discrepancy across domains. CORAL and MMD are widely acknowledged as prominent explicit distribution discrepancy criteria due to their stability and reliability, surpassing implicit alternatives. CORAL's main goal is to adjust covariance across various dimensions in the original sample space. Despite the numerous applications of MMD-based techniques in fault diagnosis, these approaches compute the disparity in mean statistics between both domains within a high-dimensional Hilbert space. In contrast to CORAL, the MMD criteria demonstrate a greater capacity for representing discrepancies as they enhance the division of data in high-dimensional space based on pattern recognition principles. Vibration signals serve as a primary means of identifying the healthy states of mechanical equipment. These signals typically exhibit symmetry along the x=0 axis and follow a Gaussian distribution. Therefore, mean statistics-based MMD may perform poorly in representing disparities under certain circumstances \cite{gilo2024subdomain}. According to \cite{qian2023maximum}, the discrepancy representation of the mean square value demonstrates superior performance compared to the mean representation. Additionally, the probability density functions (PDFs) of different domain distributions exhibit significant overlap when considering the statistical mean. As a biased two-order statistic, the mean square value is well-suited for depicting gaps in vibration signals since it can fully express the mean and variance information. However, similar to CORAL, the performance of domain confusion is significantly constrained by the low-dimensional sample space. Thus, developing a novel approach to address this challenge is imperative.

\par The reliance on labeled data representing all classes within the source domain poses a significant challenge for these methods. However, in many applications, only data representing the healthy state is available, limiting the effectiveness of these approaches. To overcome this limitation, using synthetic data to improve fault diagnosis performance is a promising approach \cite{NI2023110544,HA2023110680}. For example, synthetic vibration data can be generated using physical models that consider working conditions and bearing parameters as input, thereby enhancing the accuracy of bearing fault diagnosis \cite{brito2023fault}. The use of synthetic data generated from various faults allows for effective training and evaluation of data-driven methods using real industry data. However, this approach has several notable limitations. Firstly, to create an accurate and realistic physical model of bearing faults, comprehensive and precise information about the operating conditions of the rotating machinery is required. Additionally, the activities and noise generated by other equipment can introduce changes to the specified working conditions in the model. Secondly, even with advanced synthetic datasets, there is always a distribution discrepancy between synthetic and actual data. When the model is trained solely on synthetic data and evaluated using actual data, this distribution gap leads to significant performance degradation. Thirdly, while synthetic data generated by physical models offers benefits, it does not utilize existing unlabeled data that may contain valuable information about different faults. Lastly, constructing precise physical models for each bearing in practical applications is time-consuming and costly. Furthermore, due to the complexity of modern systems and the degradation process from normal to failure stages, it is impractical to design physics-based models and generate sufficient synthetic data for specific systems \cite{wang2021integrating}. To address these limitations, DA methods have proven effective. These strategies have been widely utilized recently to mitigate the distribution disparity caused by changing operational circumstances in real data. DA approaches seek to reduce the distribution gap between domains by incorporating synthetic data from the source domain and unlabeled real data from the target domain. This strategy eliminates the need for highly precise physical models and enables the extraction of meaningful information from unlabeled real-world data.

\begin{figure}
\label{imbalanced}
    \centering
    \includegraphics[width=\columnwidth]{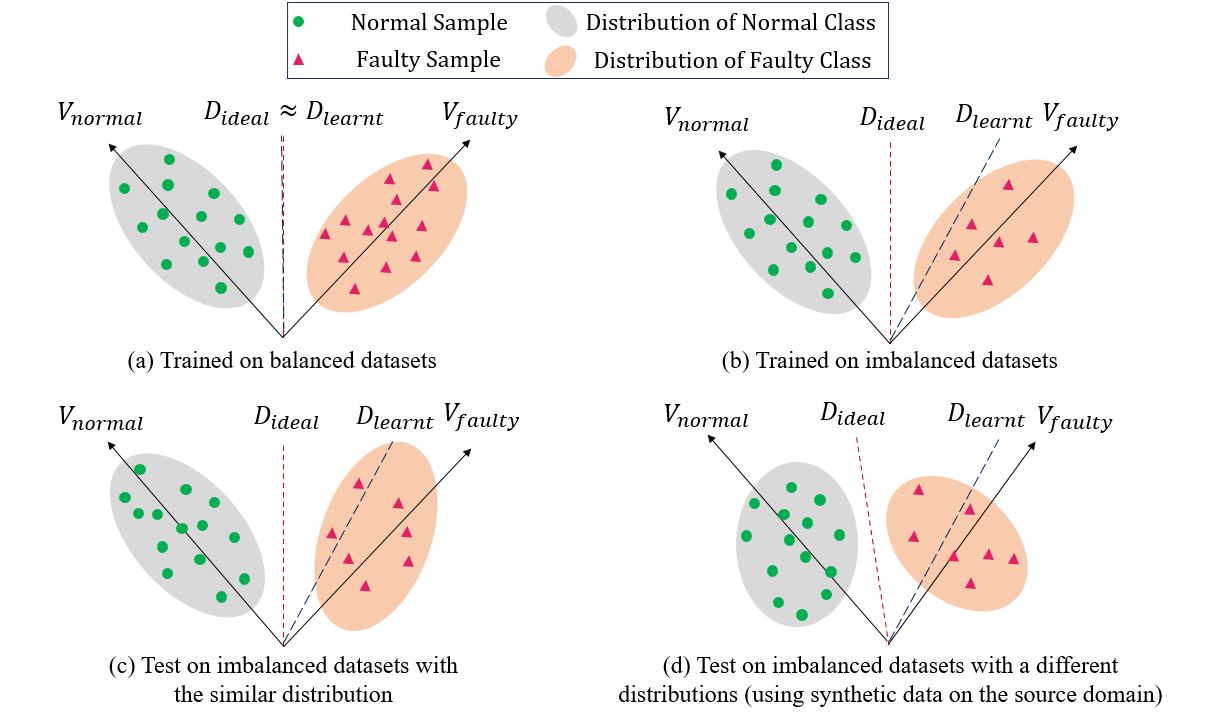}
    \caption{The effect of class-imbalance and distribution discrepancy in the fault diagnosis task. The feature vectors $V_{normal}$ and $V_{faulty}$ correspond to the learned feature representations of the normal and faulty classes, respectively. $D_{ideal}$ denotes the ideal decision boundary of the classifier, while $D_{learnt}$ represents the actual decision boundary learned from the provided datasets.}
\label{imbalanced}
\end{figure}

\par All the research described presumes that the data from various classes are structurally complete. The phenomena of missing data may occur in the real dataset due to sensor failure, random data packet dropout, deactivated monitors, and database error \cite{kavianpour2022class}. In this case, all or a portion of the data from one or more classes may be entirely missed, or some data from one or more classes may be lost. The absence of data for a specific class or classes within the target domain results in the target label space becoming a subset of the source label space. This scenario is commonly encountered in practical industrial applications. This type of problem is called partial-set fault diagnosis (PSFD) \cite{wang2023novel}. Also, when a portion of the samples from one or more classes are missed, a data imbalanced challenge emerges \cite{ren2023systematic}.
\par The DA approaches were developed for closed-set fault diagnosis (CSFD) problems with the same domain label space. A substantial class-level mismatch arises if these approaches are applied to the challenges of PSFD. The absence of labels on the target samples renders the identification of outlier classes within the source domain infeasible. Consequently, the primary obstacle in the PSFD is identifying domain invariant features for knowledge transfer and constraining source-only-class samples to prevent negative transfer \cite{XIA2024107848}. Specifically, source-only-class samples degrade positive transfer by aligning classes in the target domain with outlier classes in the source domain. This issue demonstrates that the PSFD scenario is more complicated than the CSFD. Therefore, new strategies are required to match source-shared classes in two domains with TL methods. The work presented in \cite{9787391} employs a method that focuses on aligning the distributions of shared classes across domains using a class-level weighting technique. In \cite{9763065}, a combination of multi-scale weighting and adversarial domain adaptation methods are employed to alleviate the impact of outlier classes present in the source domain. In \cite{zhu2023partial}, an innovative approach for PSFD is introduced, leveraging adversarial learning, an enhanced convolutional block attention module, and a weighting mechanism. However, the efficacy of these methods diminishes when confronted with the challenge of imbalanced datasets.

\par The domains are considered to be class-wise balanced in DA methods. The nature of many faults makes their occurrence uncommon. As a result of using the DA approaches, the distribution of unlabeled faulty data in the target domain in latent space is increasingly pulled towards the healthy class. Furthermore, no information is available on the distribution of the target classes or the imbalance level of its classes. This issue exists exclusively in the target domain. In contrast, in the source domain, samples of faulty classes are generated based on healthy class samples, resulting in a class-wise balanced source domain. As a result, the performance of conventional DA approaches for imbalanced datasets experiences a significant decline, as illustrated in Figure \ref{imbalanced}. The presence of imbalanced class introduces biases in the decision boundary of the classifier, particularly affecting minority classes. Additionally, the distribution difference between the domains leads to variations, further diminishing the classifier's ability to generalize \cite{TIAN2024109832,LEE2024122910}. When a fault diagnosis method trained on the source domain is employed to infer target datasets, the overlapping conditions aggravate its unreliability. As a solution to this problem, many regularization approaches, such as Mixup \cite{zhang2017mixup}, Manifold Mixup \cite{verma2019manifold}, and CutMix \cite{yun2019cutmix} have been presented in various disciplines. The fundamental principle behind these techniques involves training deep learning models using mixed samples generated through a convex combination of input-label or input-pseudo-label pairs. Notably, Mixup-based approaches employ linear interpolations of feature vectors alongside corresponding label interpolations, utilizing the same mixing factor. However, this approach proves ineffective when confronted with highly imbalanced class distributions.

\par This study presents a new Partial Transfer learning based on the Physics-informed Artificial Intelligence (PTPAI) method to address unsupervised fault diagnosis in scenarios where labeled faulty data is unavailable and missing data is present. Motivated by the mean square statistic and the relationship as mentioned earlier, a novel distribution discrepancy metric called the multi-kernel maximum mean square discrepancy (MK-MMSD) metric is proposed. This metric leverages the properties of the kernel function within the reproducing kernel Hilbert space (RKHS). Additionally, the CDAN technique is employed to align the distributions conditioned on classes, using pseudo-labels to provide class knowledge to the discriminator. By using MK-MMSD and CDAN, the proposed method effectively decreases the distributional discrepancy between synthetic data generated in the source domain and actual data in the target domain. A self-adaptive multi-weight block is employed to tackle the PSFD problem arising from missing data, which has three modules: class-level weighting, source-instance weighting, and target-instance weighting. Class-level weighting is applied to distinguish the label space, with the resulting weights used to modify the loss function. An auxiliary classifier, based on the source domain, and an auxiliary domain discriminator operating in the feature space are employed to quantify target samples and assess the transferability of source samples, respectively.
In the feature space, the novel re-balanced feature Mixup (RF-Mixup) regularization technique addresses the challenge of an imbalanced target dataset caused by missing data. To mitigate the impact of an imbalanced dataset, RF-Mixup can augment features and pseudo-labels and provide them as input features to other blocks. Unlike other Mixup-based approaches, RF-Mixup pushes the decision boundaries towards majority classes by considering various mixing factors for features and pseudo-labels. Consequently, it improves generalizability and decreases the impact of data imbalance.
The suggested PTPAI approach is assessed with two datasets, CWRU and JNU. Simulation results demonstrate its superior performance compared to other approaches under various missing data scenarios. While originally designed for synthetic data issues, the PTPAI technique can also be used for real-world data problems with distributional differences between domains. This includes addressing issues related to changing working conditions in rotary machinery.
\par The primary accomplishment of the current work is:
\begin{enumerate}

    \item In the source domain, synthetic data encompassing various bearing health conditions have been generated using a straightforward procedure for synthetic data generation. Furthermore, the real data in the target domain was found to contain missing data, which was incorporated to align with real-world applications. To the best of our knowledge, this is the first paper in which this particular issue has been addressed.
    \item A novel MK-MMSD technique is introduced to address the PSFD problem and mitigate the imbalanced issue. This technique effectively utilizes the properties of the kernel function in the RKHS to comprehensively capture and represent the mean and variance information of data samples.
   \item A novel approach called RF-Mixup is presented to augment the distribution, ensuring the proposed method remains robust to varying levels of imbalanced class.
   \item To tackle the PSFD problem, a novel fault diagnosis approach is proposed, which combines MK-MMSD and CDAN techniques. This approach attempts to mitigate the distribution disparity between synthetic data in the source domain and actual data in the target domain by utilizing weighting blocks.

\end{enumerate}

\section{preliminariess}
\subsection{ Problem formulation}
This paper addresses the problem of partial imbalanced transfer learning within the context of bearing fault diagnosis. The approach assumes the use of synthetic labeled data in the source domain, while real unlabeled data is utilized in the target domain.
Let ${\mathscr{D}_s} = \left\{ {\left( {{x_{s,i}},{y_{s,i}}} \right)} \right\}_{i = 1}^{{n_s}}$ and ${\mathscr{D}_t} = \left\{ {\left( {{x_{t,j}}} \right)} \right\}_{j = 1}^{n_t}$ indicate the source domain with ${{n_s}}$ labeled samples, and the target domain with ${n_t}$ unlabeled samples, respectively. Here, ${{x_{s,i}}}$ and ${{x_{t,j}}}$ denote features of the source and target domains, respectively, while ${{y_{s,i}}}$ represents the label space of the source domain. It is assumed that ${\mathscr{D}_s}$ and ${\mathscr{D}_t}$ are drawn from the distributions ${p_s}$ and ${p_t}$, respectively. In partial transfer learning, the target domain's classes ${\mathcal{C}_t}$ are a subset of the source domain's classes ${\mathcal{C}_s}$, i.e., ${\mathcal{C}_t} \subset {\mathcal{C}_s}$.
Additionally, the quantity of faulty classes samples present in the target domain is lower in comparison to the analogous class samples within the source domain. Hence, in the target domain, within the target domain, the predominant class comprises healthy instances, whereas each defective class constitutes a minority. Synthetic data in the source domain and actual data in the target domain contribute to a distribution gap across domains. As a result of the domain shift phenomena, ${p_s} \ne {p_t}$. Also, it is presumed that ${p_{s,shared}} \ne {p_t}$ where ${p_{s,shared}}$ represents the distribution of the source domain in the shared space of domains.

\subsection{Maximum mean square discrepancy}

The concept of MMD entails assessing the average disparities between the mappings of two domains within the RKHS. The MMD metric is quantifiable through the following equation:

\begin{dmath}
    $${MMD}[{\mathcal{H}},{\mathscr{D}_s},{\mathscr{D}_t}] = \mathop {sup}\limits_{{{\left\| f \right\|}_{\mathcal{H}}} \leqslant 1} \left| {{E_{{p_s}}}\left[ {f({x_s})} \right] - {E_{{p_t}}}\left[ {f({x_t})} \right]} \right| = \mathop {sup}\limits_{{{\left\| f \right\|}_{\mathcal{H}}} \leqslant 1} \left| {{{\left\langle {f,{u_{{p_s}}}} \right\rangle }_H} - {{\left\langle {f,{u_{{p_t}}}} \right\rangle }_{\mathcal{H}}}} \right| = \mathop {sup}\limits_{{{\left\| f \right\|}_{\mathcal{H}}} \leqslant 1} \left( {{{\left\langle {f,{u_{{p_s}}} - {u_{{p_t}}}} \right\rangle }_{\mathcal{H}}}} \right) \leqslant {\left\| f \right\|_{\mathcal{H}}}{\left\| {{u_{{p_s}}} - {u_{{p_t}}}} \right\|_{\mathcal{H}}} \\ \leqslant {\left\| {{u_{{p_s}}} - {u_{{p_t}}}} \right\|_{\mathcal{H}}}$$
    \label{mmd1}
\end{dmath}

where $<.,.>$ is the inner product, $E$ is expected value, $\mathcal{H}$ indicates a Hilbert space, and $f \in \mathcal{H}$. The condition ${{\left\| f \right\|_H} \leqslant 1}$ signifies the utilization of the unit RKHS. ${u_{{p_s}}}$ and ${u_{{p_t}}}$ refer to the mean embedding of the probability distributions in the source and target domains, respectively. The mathematical expectation of the sample mappings from the two domains can be calculated in the following manner:
\begin{dmath}
    $${E_{{p_s}}}\left[ {f({x_s})} \right] = \int_{{x_s} \in {\mathscr{D}_s}} {{p_s}({x_s})} f({x_s})d{x_s} = \int_{{x_s} \in {\mathscr{D}_s}} {{p_s}({x_s})} {\left\langle {k({x_s},),f} \right\rangle _{\mathcal{H}}}d{x_s} = {\left\langle {f,\int_{{x_s} \in {\mathscr{D}_s}} {{p_s}({x_s})k({x_s},)} d{x_s}} \right\rangle _{\mathcal{H}}} = {\left\langle {f,{u_{{p_s}}}} \right\rangle _{\mathcal{H}}}$$
    \label{mmd2}
\end{dmath}

\begin{dmath}
   $${E_{{p_t}}}\left[ {f({x_t})} \right] = \int_{{x_t} \in {\mathscr{D}_t}} {{p_t}({x_t})} f({x_t})d{x_t} = \int_{{x_t} \in {\mathscr{D}_t}} {{p_t}({x_t})} {\left\langle {k({x_t},),f} \right\rangle _{\mathcal{H}}}d{x_t} = {\left\langle {f,\int_{{x_t} \in {D_t}} {{p_t}({x_t})k({x_t},)} d{x_t}} \right\rangle _{\mathcal{H}}} = {\left\langle {f,{u_{{p_t}}}} \right\rangle _{\mathcal{H}}}$$ 
   \label{mmd3}
\end{dmath}

where $k(x,.)$ is kernel function and is utilized to create a Hilbert space $\mathcal{H} = span\{ k(x,.)\left| {x \in X} \right.\} $. The relation between the illustration of differences in MMD and the kernel function is shown by Eq. \ref{mmd1} to Eq. \ref{mmd3}. Based on the examination presented in Section \ref{sec1}, it is evident that the use of the mean square statistic is more appropriate for capturing the disparity in distribution across data samples obtained from vibration signals, as compared to the mean statistic. However, the ability to describe discrepancies in a low-dimensional sample space is restricted compared to a high-dimensional space. The operational process of MMD reveals that the kernel function $k({x_s},.)$ serves as the sample mapping inside an infinite-dimensional RKHS. The measurement of the distribution gap between the mean square statistics of both domains can be conveniently conducted using the tensor product of the kernel function $k({x_s},.)$ and itself, denoted as ${k({x_s},.) \otimes k({x_s},.)}$ \cite{qian2023maximum}. MMSD can be computed using the following equations:

\begin{equation}
    \begin{split}
        & MMSD[\mathcal{H} \otimes \mathcal{H},{\mathscr{D}_s},{\mathscr{D}_t}] =\\ 
        & \quad\mathop{\sup}\limits_{{\| h \|}_{\mathcal{H}_1 \otimes \mathcal{H}_2} \leqslant 1} \left| \left\langle f,{E_{{p_s}}}\left[ k({x_s},.) \otimes k({x_s},.) \right] \right\rangle_{\mathcal{H} \otimes \mathcal{H}} - \right.\\
        & \quad\quad\left. \left\langle f,{E_{{p_t}}}\left[ k(.,{x_t}) \otimes k(.,{x_t}) \right] \right\rangle_{\mathcal{H}\otimes \mathcal{H}} \right|
    \end{split}
\end{equation}

\begin{equation}
    \begin{multlined}
        $$MMSD[\mathcal{H} \otimes \mathcal{H},{\mathscr{D}_s},{\mathscr{D}_t}] \leqslant \\{\left\| f \right\|_{\mathcal{H} \otimes \mathcal{H}}}{\left\| {{E_{{p_s}}}\left[ {k({x_s},.) \otimes k({x_s},.)} \right]  - {E_{{p_t}}}\left[ {k(.,{x_t}) \otimes k(.,{x_t})} \right]} \right\|_{H \otimes \mathcal{H}}}\\ \leqslant {\left\| {{E_{{p_s}}}\left[ {k({x_s},.) \otimes k({x_s},.)} \right] -  {E_{{p_t}}}\left[ {k(.,{x_t}) \otimes k(.,{x_t})} \right]} \right\|_{\mathcal{H}\otimes \mathcal{H}}}$$
    \end{multlined}
    \label{mmsd}
\end{equation}

The tensor product of two Hilbert spaces remains a Hilbert space itself. In Eq. \ref{mmsd}, the tensor product $k(x,.) \otimes k(x,.)$ is employed as a substitute for the inner product $k(x,.) \cdot k(x,.)$ in order to preserve the operating features of MMD and decrease the computational cost.

\begin{dmath}
    $$\left\langle {k({x_s},.) \otimes k({x_s},.),k(.,{x_t}) \otimes k(.,{x_t})} \right\rangle  = {\left\langle {k({x_s},.),k(.,{x_t})} \right\rangle ^2}$$
\end{dmath}

The Equation used for calculating the final square MMSD is formally stated as follows:


\begin{dmath}
        $$MMS{D^2}[\mathcal{H} \otimes \mathcal{H},{\mathscr{D}_s},{\mathscr{D}_t}] = \\ \left\| {{E_{{p_s}}}\left[ {{k}({x_s},.) \otimes k({x_s},.)} \right] - \\{E_{{p_t}}}\left[ {k(.,{x_t}) \otimes k(.,{x_t})} \right]} \right\|_{\mathcal{H} \otimes \mathcal{H}}^2$$ 
        \label{mmsd2}
\end{dmath}

For using the MMSD criteria in Eq. \ref{mmsd2} to real-world scenarios, the biased empirical MMSD statistic $MMSD_b^2[H \otimes H,{\mathscr{D}_s},{\mathscr{D}_t}]$ is used as follows:

\begin{dmath}
\label{MMSD3}
    $$MMSD_b^2[\mathcal{H} \otimes \mathcal{H},{\mathscr{D}_s},{\mathscr{D}_t}] = \frac{1}{{n_s^2}}\sum\limits_{i = 1}^{{n_s}} {\sum\limits_{j = 1}^{{n_s}} {{{\left\langle {k(x_i^s,.),k(x_j^s,.)} \right\rangle }^2}}  + } \frac{1}{{n_t^2}}\sum\limits_{i = 1}^{{n_t}} {\sum\limits_{j = 1}^{{n_t}} {{{\left\langle {k(.,x_i^t),k(.,x_j^t)} \right\rangle }^2}} }  - \frac{2}{{{n_s}{n_t}}}\sum\limits_{i = 1}^{{n_s}} {\sum\limits_{j = 1}^{{n_t}} {{{\left\langle {k(x_i^s,.),k(.,x_j^t)} \right\rangle }^2}} }  = \frac{1}{{n_s^2}}\sum\limits_{i = 1}^{{n_s}} {\sum\limits_{j = 1}^{{n_s}} {{k^2}(x_i^s,x_j^s)}  + } \frac{1}{{n_t^2}}\sum\limits_{i = 1}^{{n_t}} {\sum\limits_{j = 1}^{{n_t}} {{k^2}(x_i^t,x_j^t)} }  - \frac{2}{{{n_s}{n_t}}}\sum\limits_{i = 1}^{{n_s}} {\sum\limits_{j = 1}^{{n_t}} {{k^2}(x_i^s,x_j^t)} } $$
\end{dmath}

\subsection{Synthetic data generation}
One of the most fundamental obstacles to real-world bearing fault diagnosis is the lack of access to realistic faulty data. The physics-informed deep learning approach by producing synthetic data can effectively solve this issue. While synthetic data attempts to mitigate the lack of labeled data, a domain shift exists between synthetic and realistic faulty data. This shift arises from factors such as inaccurate modeling and environmental variations. The performance of deep learning algorithms may drop significantly due to this domain shift. Furthermore, the absence of access to actual rare faulty data labels significantly limits the generalization capability.

\par One of the difficulties in employing physics-informed AI approaches is creating synthetic data without using very precise physical models. We can presume that we will have access to some real healthy data in order to generate synthetic data. In practical applications, it is a reasonable assumption that healthy data can be acquired by recording early from rotary machinery before faults occur. Healthy real data are used to generate faulty data as well as encode knowledge about working conditions and environmental noise into synthetic faulty data.
According to \cite{randall2001relationship,borghesani2013new}, a common method is employed to obtain synthetic faulty bearing data. A modeled vibration signal $\xi$ at time $t$ can be described as follows:
\begin{equation}
\label{synthetic}
        \xi (t) = \sum\limits_{j =  - \infty }^{ + \infty } {{\mathcal{A}_j}s(t - jT) + \beta n(t)} 
\end{equation}
where $s$ is the oscillating waveform triggered by a single impact caused by over-rotation of a surface fault with period $T$.
${\mathcal{A}_j}$ with period $Q$ indicates the amplitude of the $i^{th}$ impact and is used to replicate a modulation term caused by an inner ring or a rolling element fault.
The parameters $T$ and $Q$ can be derived from the bearing's kinematics and a speed recorded by a domain expert \cite{randall2011rolling}.
The term $n$ is considered the additive ambient noise equivalent to the vibration recording of a healthy bearing.
$\beta$ is uniformly distributed between 0.25 and 2 to guarantee that this approach is robust over a broad range of signal-to-noise ratios (SNRs). A Hann window with a duty period of 5\% concerning $T$ is used to depict a single impact $s$. Due to the lack of information from the transfer function, a wide-band band-pass filter is applied to $s$. This filter decreases frequencies near 0 $Hz$ and the Nyquist frequency primarily. Since fault frequencies are visible in all frequency bands, the DA approach can effectively reduce distribution discrepancy across actual and synthetic data. $\mathcal{A}_j$ is defined as follows: 
\begin{equation}
        {\mathcal{A}_j} = \lambda \sum\limits_{m = 0}^M {{\alpha _m}\cos (\frac{{2m\pi Tj}}{Q})} 
\end{equation}
where $m$ indicates the number of sidebands and $m$ can be
controlled with associated amplitudes $\alpha$ in the envelope
analysis. For adding some natural randomness to the impacts
$s$, $\lambda$ is used. $\lambda$ is normally distributed with
a mean of 1 and a standard deviation of 0.1.

\subsection{Mixup regularization technique}
Mixup training is a practical approach for creating new training data by convex combining input data. This approach successfully enhances model generalization while reducing training data memory. 
Mixup training has recently been used in DA methods \cite{xu2020adversarial,ma2023source}. Mixup improves domain invariance and can significantly increase the cross-domain model's generalization capabilities. 
Mixup training can be calculated as follows: 
\begin{equation}
    \label{mixup}
    \left\{ {\begin{array}{*{20}{c}}
  {x = \lambda {x_i} + (1 - \lambda ){x_j}} \\ 
  {y = \lambda {y_i} + (1 - \lambda ){y_j}} 
\end{array}} \right.
\end{equation}
where $\lambda$ is obtained from a prior $\beta$ distribution, with $\lambda  \sim \beta ({\alpha _1},{\alpha _2})$, in which ${\alpha _1}$ and ${\alpha _2}$ regulate the shape of the $\beta$ distribution. The typical mixup training method usually interpolates linearly between samples and labels while utilizing the identical mixing factor $\lambda$. For an imbalanced dataset, the produced samples and labels are still skewed towards the class with the greater sample amounts. The model may then concentrate on the samples that contribute most to the loss function.

\section{Proposed method}
\subsection{Framework of the proposed method}

\begin{figure*}[th]
    \centering
    
    \includegraphics[width=\textwidth]{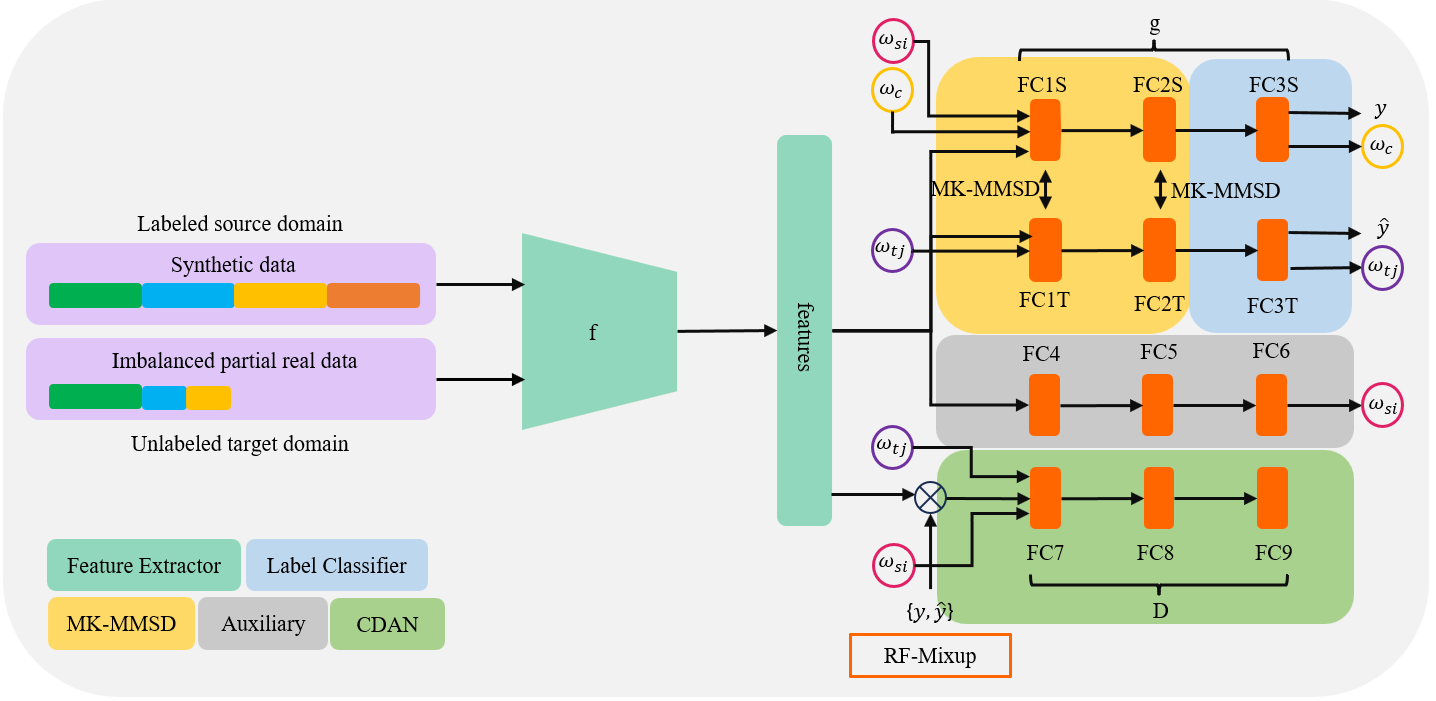}
    \caption{Overall framework of the proposed PTPAI method}
    \label{proposed}
\end{figure*}

This section presents the PTPAI method, which is proposed for fault diagnosis when dealing with incomplete information in certain instances and classes, along with a strategy to mitigate the distribution difference between synthetic and real data. The PTPAI method is composed of five blocks, as illustrated in Figure \ref{proposed}. These blocks include the input block, the feature extractor block, the RF-mixup block, the weighting block, and the transfer learning block. Also, based on different objective functions and feature extractor, five parts of the proposed PTPAI method  is shown in this figure.  The input block contains synthetic source domain data and real target domain data. The feature extractor block employs CNN to capture spatial characteristics from the input data. The RF-mixup block generates features by combining the features from the previous block with soft labels, which are then utilized as input for the subsequent blocks. This block addresses the challenge of imbalanced datasets, aiming to prevent optimization instability and provide more data for the training phase of the proposed model. The weighting block incorporates class-level and instance-level weighting, which helps tackle outlier classes in the source domain by determining coefficients for modules in the transfer learning block. The transfer learning block, consisting of MK-MMSD and CDAN modules, aims to bridge the distribution gap between synthetic and real domain data. Table \ref{architecture} outlines the structural details of the proposed method. Each block will be elaborated upon in detail, culminating in an explanation of the objective function and its optimization at the end of this section.

\subsubsection{Input block}
This block consists of labeled synthetic data generated in the source domain using Eq. \ref{synthetic} and unlabeled real data collected in the target domain. Due to the generation of faults based on healthy class data in the source domain, each class possesses an equal number of samples, resulting in a balanced source domain. Conversely, the target domain exhibits an imbalanced class distribution, with a healthy class containing the same number of samples as the source domain's healthy class and some faulty classes with limited samples. Moreover, only a subset of classes shared between the source and target domains exists, leading to the presence of outlier classes in the source domain. This block utilizes the full-wave rectified envelope spectrum of the data in both domains as input.
\subsubsection{Feature extractor block} The aforementioned block comprises multiple convolution layers aimed at extracting intricate features at a higher level. Through the utilization of convolutional layers and employing various techniques, this block adeptly captures spatial characteristics from both synthetic and real data. The abstract features obtained from this process subsequently serve as valuable inputs for subsequent blocks in the architecture \cite{kavianpour2023cnnlstmAM}.

\subsubsection{RF-Mixup block}
\begin{figure}[t]

    \centering
    \includegraphics[width=\columnwidth]{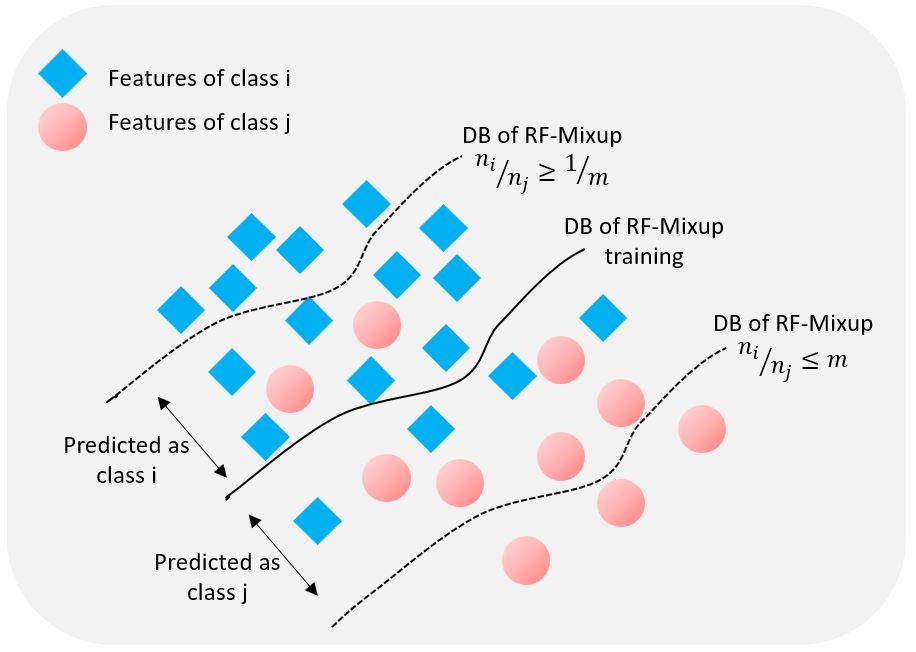}
    \caption{The comparison of the RF-mixup with several DBs. With $m<1$, the RF-mixup allocates the label from the minority class to a greater value. When $m>1$, RF-mixup produces labels that support the majority class. As a result, this paper only considers $m<1$.}
    \label{RF-MIXUP}
    \end{figure}

The limited availability and lack of labels for faulty data in the target domain present a significant challenge for cross-domain fault diagnosis. The imbalanced nature of the dataset, attributed to a limited number of faulty samples, can introduce instability in optimizing the loss function and adversely impact domain alignment performance. To tackle this challenge, we propose an innovative approach named RF-Mixup, which builds upon the Mixup technique. By leveraging augmented features and pseudo-labels derived from the unlabeled imbalanced datasets in the target domain, RF-Mixup enhances the input for the CDAN module. Given a specific batch of input data $x$, we can compute the feature embedding vector $e=f(x)$ and obtain the corresponding pseudo-label vector $\overset{\lower0.5em\hbox{$\smash{\scriptscriptstyle\frown}$}}{y}=g(f(x))$. These feature embedding and pseudo-label vectors are derived from the actual target dataset. The calculation of RF-Mixup is performed as follows:


\begin{equation}
\label{erfm}
    {e^{RFM}} = {\lambda _e}{e_i} + (1 - {\lambda _e}){e_j}
\end{equation}

\begin{equation}
\label{yrfm}
   {y^{RFM}} = {\lambda _y}{{\overset{\lower0.5em\hbox{$\smash{\scriptscriptstyle\frown}$}}{y} }_i} + (1 - {\lambda _y}){{\overset{\lower0.5em\hbox{$\smash{\scriptscriptstyle\frown}$}}{y} }_j} 
\end{equation}

\begin{equation}
    {z^{RFM}} = {e^{RFM}} \otimes {y^{RFM}}
\end{equation}
where ${\lambda _e}$ and ${\lambda _y}$ represent the mixing factors for features and labels, respectively. The interpolated features are denoted by ${e^{RFM}}$, their class label by ${y^{RFM}}$, and the multi-linear input to the CDAN module by ${z^{RFM}}$. $\otimes$ is a multi-linear map.
${\lambda _e}$ is determined randomly from a $\beta$ distribution as in Eq. \ref{mixup}. In the case of ${\lambda _y}$, the RF-mixup approach provides a decision boundary and substitutes the real labels with labels that have a larger value for the minority class. ${\lambda _y}$ is formulated as:
\begin{equation}
\label{lambday}
{\lambda _y} = \left\{ {\begin{array}{*{20}{c}}
  \begin{gathered}
  \max ({\lambda _e},1 - {\lambda _e}) \hfill \\
  \min ({\lambda _e},1 - {\lambda _e}) \hfill \\
  {\lambda _e} \hfill \\ 
\end{gathered} &\begin{gathered}
  {{n_i} \mathord{\left/
 {\vphantom {{n_i} {n_j \leqslant m}}} \right.
 \kern-\nulldelimiterspace} {n_j \leqslant m}} \hfill \\
  {{n_i} \mathord{\left/
 {\vphantom {{n_i} {n_j \geqslant {1 \mathord{\left/
 {\vphantom {1 m}} \right.
 \kern-\nulldelimiterspace} m}}}} \right.
 \kern-\nulldelimiterspace} {n_j \geqslant {1 \mathord{\left/
 {\vphantom {1 m}} \right.
 \kern-\nulldelimiterspace} m}}} \hfill \\
  otherwise \hfill \\ 
\end{gathered}  
\end{array}} \right.
\end{equation}
where ${n_i}$ and ${n_j}$ are the number of samples in classes $i$ and $j$ in the target domain, respectively. The decision boundary (DB) for label reassignment is controlled by $m$.
Figure \ref{RF-MIXUP} depicts the RF-mixup procedure and its comparison with the Mixup.

\par Assume that ${e_i}$ has 1200 samples from class 1 and ${e_j}$ has 24 samples from class 2. When ${\lambda _e} = 0.8$, the label is set via the mixup training to be 80\% of class 1 and 20\% of class 2. RF-mixup allocates the soft label to be 20\% class 1 and 80\% class 2, as shown in Eq. \ref{lambday}, where ${{n_i^t} \mathord{\left/
 {\vphantom {{n_i^t} {n_j^t \geqslant {1 \mathord{\left/
 {\vphantom {1 m}} \right.
 \kern-\nulldelimiterspace} m}}}} \right.
 \kern-\nulldelimiterspace} {n_j^t \geqslant {1 \mathord{\left/
 {\vphantom {1 m}} \right.
 \kern-\nulldelimiterspace} m}}}$ for $m = 0.5$. RF-mixup enhances data diversity and enables the proposed method to focus on minority class data through the reassignment of soft labels that favor the minority class. In this work, we utilize these generated pseudo-labels during the training phase to improve the model's performance on imbalanced data.

\begin{table*}[th]
\caption{Network architecture of the PTPAI method}
\label{architecture}
\resizebox{\textwidth}{!}{%
\begin{tabular}{cccccc}
\hline
Parts &
  Layers &
  Size/stride/number of kernels or number of neurons &
  Pooling size &
  Padding &
  Dropout \\ \hline
\multirow{4}{*}{Feature Extractor} &
  Conv1, BN, ReLU, Max-pooling &
  128/2/16 &
  2 &
  \checkmark  &
  - \\
                           & Conv2, BN, ReLU, Max-pooling & 64/2/32            & 2 & \checkmark  & -   \\
                           & Conv3, BN, ReLU, Max-pooling & 3/2/64             & 2 & \checkmark  & -   \\
                           & Conv4, BN, ReLU, GAP         & 3/2/128            & 2 & \checkmark  & -   \\ \hline
\multirow{2}{*}{MK-MMSD}   & FC1, ReLU, Dropout           & 256                & - & -                         & 0.5 \\
                           & FC2, ReLU, Dropout           & 128                & - & -                         & 0.5 \\ \hline
Label Classifier           & FC3, Softmax                 & Number of category & - & -                         & -   \\ \hline
\multirow{3}{*}{Auxiliary} & FC4, ReLU, Dropout           & 256                & - & -                         & 0.5 \\
                           & FC5, ReLU, Dropout           & 128                & - & -                         & 0.5 \\
                           & FC6, Leaky Softmax           & Number of category & - & -                         & -   \\ \hline
\multirow{3}{*}{CDAN}      & FC7, ReLU, Dropout           & 256                & - & -                         & 0.5 \\
                           & FC8, ReLU, Dropout           & 128                & - & -                         & 0.5 \\
                           & FC9, Softmax                 & 2                  & - & -                         & -   \\ \hline 
\end{tabular}%
}
\end{table*}

\subsubsection{Weighting block}
Conventional DA methods operate under the assumption that the source and target domains share identical class labels. However, the presence of outlier classes in the source domain can negatively impact the performance of DA methods. While addressing partial domain challenges using conventional DA methods is feasible through the identification and removal of outlier samples in the source domain, the unlabeled nature of the target domain presents a significant obstacle in determining the outlier class(es) in the source domain. Several weight functions are employed to overcome this challenge. These weight functions are incorporated into the loss function to assign different weights to the impact of various classes and instances and decrease the impact of outlier source classes. Applying these weights minimizes the impact of outlier classes, thereby mitigating their negative effect on the domain adaptation process.


\par 4-1- \textbf{Class-Level Weighting:} The classifier's output for each instance ${x_i}$ indicates the probability distribution in the labeled space of the source domain. It determines the probability of allocating ${x_i}$ to each $\left| {{\mathcal{C}_s}} \right|$ classes. Given the disjoint nature of the outlier label space in the source domain and the label space in the target domain, the target data is expected to exhibit significant differences from the outlier classes. Consequently, the probability of assigning target data to outlier classes should be minimal. To address this, class-level weighting has been employed to evaluate the significance of applying the classification loss function and to emphasize the importance of shared and outlier classes during the training process. The classifier trained on the source domain data is then applied to instances in the target domain, resulting in the following:

\begin{equation}
    \begin{aligned}
        {{\tilde Y}^t} = \left[ {\begin{array}{*{20}{c}}
  {\tilde y_1^t} \\ 
  {\tilde y_2^t} \\ 
   \vdots  \\ 
  {\tilde y_{{n_t}}^t} 
\end{array}} \right] = \left[ {\begin{array}{*{20}{c}}
  {\tilde y_{1,{c_1}}^t}&{\tilde y_{1,{c_2}}^t}& \cdots &{\tilde y_{1,{c_{\left| {{C_s}} \right|}}}^t} \\ 
  {\tilde y_{2,{c_1}}^t}&{\tilde y_{2,{c_2}}^t}& \cdots &{\tilde y_{2,{c_{\left| {{C_s}} \right|}}}^t} \\ 
   \vdots & \vdots & \ddots & \vdots  \\ 
  {\tilde y_{{n_t},{c_1}}^t}&{\tilde y_{{n_t},{c_2}}^t}& \cdots &{\tilde y_{{n_t},{c_{\left| {{C_s}} \right|}}}^t} 
\end{array}} \right]
    \end{aligned}
\end{equation}
where $\tilde y_i^t$ denotes the probability of instance $x_i^t$ belonging to each ${{\mathcal{C}_s}}$ classes and $i=1,2,...,{{n_t}}$. To mitigate the effects of randomness and infrequent errors, the average of the predicted labels for all target data is used as the class-level weight, as shown below:
\begin{equation}
\label{wc}
    \begin{aligned}
        {\omega _c} = \frac{1}{{{n_t}}}\sum\limits_{i = 1}^{{n_t}} {\tilde y_i^t} 
    \end{aligned}
\end{equation}
where ${\omega _c}$ is a vector of ${{\mathcal{C}_s}}$ dimensions. We use ${\omega _c} \leftarrow {{{\omega _c}} \mathord{\left/
 {\vphantom {{{\omega _c}} {\max ({\omega _c})}}} \right.
 \kern-\nulldelimiterspace} {\max ({\omega _c})}}$ to normalize the weight ${\omega _c}$. Due to the significant dissimilarity between the target instances and the outlier classes, the weights assigned to the outlier classes are expected to be considerably low. The outlier classes' low weight will lessen the outliers' impact and aid in more precise prediction.

\par 4-2- \textbf{Instance-level weighting}: This weighting method is employed in the source and target domains.The weighting method in each domain is described below:
\par \textbf{Source Instance Weighting}: In the proposed method, an extra auxiliary discriminator ${\mathcal{{\tilde G}}_d}$ is created to assess the significance of source instances. ${\mathcal{{\tilde G}}_d}$ is implemented to eliminate outlier samples and determine the weight function for source instances. ${\mathcal{{\tilde G}}_d}$ uses adversarial training to distinguish two domains' inputs and get the binary classification result. However, ${\mathcal{{\tilde G}}_d}$ solely focuses on domain distinction, neglecting the labeled information available in the source domain. This limitation potentially hinders the effectiveness of ${\mathcal{{\tilde G}}_d}$ as irrelevant samples may still pose challenges. To address this issue, an auxiliary classifier, ${\mathcal{{\tilde G}}_y}$, has been developed to leverage prior label information and enhance the learning process of ${\mathcal{{\tilde G}}_d}$. Unlike conventional label classifiers that utilize the softmax function, the auxiliary classifier employs a leaky-softmax activation function to translate the features extracted by the generator into an output of dimension ${\left| {{\mathcal{C}_s}} \right|}$:

\begin{equation}
    \begin{aligned}
        {{ \mathcal{\tilde{G}}}_y}(z) = \frac{{\exp (z)}}{{\left| {{\mathcal{C}_s}} \right| + \sum\limits_{c = 1}^{\left| {{\mathcal{C}_s}} \right|} {\exp ({z_c})} }}
    \end{aligned}
\end{equation}
where $z$ represents the hidden layer's output within the auxiliary classifier. Given that ${\mathcal{{\tilde G}}_d}$ is trained exclusively on labeled data from the source domain, it is reasonable to assume that ${\mathcal{{\tilde G}}_y}$ would provide output with high levels of probability and confidence for source samples while presenting lower values and less confident predictions for target domain data. The leaky-Softmax output is near one for source instances and near zero for target instances. Consequently, the output of ${\mathcal{{\tilde G}}_d}$ can be described as follows:

\begin{equation}
    \begin{aligned}
       {{\mathcal{\tilde{G}}}_d}({\mathcal{G}_f}({x_i})) = \sum\limits_{c = 1}^{\left| {{\mathcal{C}_s}} \right|} {\mathcal{\tilde G}_y^c({\mathcal{G}_f}({x_i}))} 
    \end{aligned}
\end{equation}
where $\mathcal{\tilde G}_y^c({\mathcal{G}_f}({x_i}))$ represents the probability output of each sample $x_i$ belonging to class $c$. ${{\mathcal{\tilde{G}}}_d}({\mathcal{G}_f}({x_i}))$ is defined as calculating the probability value for each sample in the source domain. During the training phase, ${{\mathcal{\tilde{G}}}_d}$ is trained to provide a discriminative output for domains. When samples from the source domain are identified as outliers, ${{\mathcal{\tilde{G}}}_d}({\mathcal{G}_f}({x_i}))$ will provide a high probability value. Conversely, a lower value of ${{\mathcal{\tilde{G}}}_d}({\mathcal{G}_f}({x_i}))$ for a source instance indicates a higher likelihood of it being identified as target domain data. This implies that the instance is closer to the target domain data and more likely to belong to the shared label space $C_t$. Consequently, the weight matrix must have an inverse relationship with ${{\mathcal{\tilde{G}}}_d}({\mathcal{G}_f}({x_i}))$, as described below:

\begin{equation}
    \begin{aligned}
       {\omega _{si}}\left( {x_i^s} \right) = 1 - {{\tilde G}_d}({G_f}({x_i}))
    \end{aligned}
\end{equation}
where the $\omega \left( {x_i^s} \right)$ represents the weight assigned to the source domain sample $x_i^s$. Notably, the weight assigned to samples belonging to outlier classes is closer to 0. The weights of each mini-batch with size $N_b$ are normalized as follows:

\begin{equation}
\label{wsi}
    {\omega _{si}}\left( {x_i^s} \right) \leftarrow \frac{{\omega _{si}\left( {x_i^s} \right)}}{{\frac{1}{{{N_b}}}\sum\limits_{i = 1}^{{N_b}} {{\omega _{si}}\left( {x_i^s} \right)} }}
\end{equation}
\par Finally, the extracted features ${\mathcal{G}_f}$ are used as input to classify the ${\left| {{\mathcal{C}_s}} \right|}$ health status classes from the source domain. This classification process trains the auxiliary discriminator ${\mathcal{\tilde{G}}}_d$ using the leaky-softmax output. The auxiliary discriminator is trained using both the labeled source domain data and the unlabeled target domain data. Subsequently, the weights of the source instances are measured. It is important to note that the auxiliary classifier is a one-to-many classifier, while the auxiliary discriminator is a binary classifier. The losses associated with the auxiliary classifier and the auxiliary discriminator are calculated as follows:

\begin{dmath}
\label{aux-class}
{\mathcal{L}_{{{\tilde G}_y}}} =  - \frac{1}{{{n_s}}}\sum\limits_{i = 1}^{{n_s}} {\sum\limits_{c = 1}^{\left| {{C_s}} \right|} {[y_{i,c}^s\log (\tilde G_y^c({G_f}(x_i^s)))} }  + (1 - y_{i,c}^s)\log (1 - \tilde G_y^c({G_f}(x_i^s)))]
\end{dmath}

\begin{dmath}
\label{aux-dis}
{\mathcal{L}_{{{\tilde G}_d}}} =  - \frac{1}{{{n_s}}}\sum\limits_{i = 1}^{{n_s}} {\log \left( {{{\tilde G}_d}({G_f}(x_i^s))} \right)}  - \frac{1}{{{n_t}}}\sum\limits_{j = 1}^{{n_t}} {\log \left( {1 - {{\tilde G}_d}({G_f}(x_j^t))} \right)} 
\end{dmath} 
where $y_{i,c}^s$ indicates whether the source instance $x_i^s$ falls within the ground-truth label of class $c$. The auxiliary discriminator's performance depends on the performance of the extracted features ${\mathcal{G}_f}$ and the auxiliary classifier ${{\tilde G}_y}$, as shown by Eq. \ref{aux-class}-\ref{aux-dis}. Consequently, it contains information about classes and domains. As a result, it can accurately determine each instance's effect, minimizing the impact of outlier classes.

\par\textbf{Target Instance Weighting}: Besides determining the weight for each instance in the source domain, it is critical to figure out the weight for each instance in the target domain. Due to the limitations of real-world industrial applications, the actual target data may include some abnormal data. As a result, a weight should be applied to each instance in the target domain to align domains and subdomains better. These weights lessen the impact of abnormal data, enhance parameter optimization and convergence, and eventually facilitate a better alignment between the domains' shared classes. To determine the weight of each instance in the target domain, it is required to consider its similarity to the source domain. Because the primary classifier of the proposed method is trained using labeled source data, it can compute the probability of each class for each target data. We assign a weight to each sample based on its maximum probability. If $x_j^t$ has the soft label $\tilde y_j^t \in {C_s}$, the instance-level weight is determined by the following:

\begin{equation}
\label{wtj}
    {\omega _{tij}} = \max (\tilde y_j^t)
\end{equation}
Moreover, all target instance weights are specified by ${\omega _{ti}} = \left\{ {{\omega _{tij}}} \right\}_{j = 1}^{{n_t}}$

\subsection{Optimization objective}
To mitigate the challenges posed by missing data, the proposed framework employs a multi-objective optimization approach during training. This approach aims to: 1) minimize classification errors on synthetic data within the source domain, 2) maximize the CDAN loss function to enhance domain adaptation in the face of missing data, 3) minimize the distribution gap between domains using the MK-MMSD metric in scenarios with missing data, and 4) minimize the losses associated with the auxiliary classifier and discriminator. Each colored rectangular box in the right side of Figure. \ref{proposed} shows an objective function.
\par \textbf{First object:} The first objective aims to minimize classification errors within the source domain. The PTPAI method leverages the cross-entropy loss function ${\mathcal{L}_C}$ to minimize the discrepancy between predicted and actual class labels. To address the potential impact of outlier classes, class-level weighting is incorporated. The ${\mathcal{L}_C}$ is calculated as follows:

\begin{dmath}
    \label{lc}
    $${\mathcal{L}_C} = \frac{1}{{{n_s}}}\sum\limits_{i = 1}^{{n_s}} {\sum\limits_{c = 1}^{\left| {{C_s}} \right|} {{w_c}{{.1}_{\left[ {y_i^s = c} \right]}}\log \frac{{\exp (z_{i,c}^{3s})}}{{\sum\limits_{j = 1}^{\left| {{C_s}} \right|} {\exp (z_{i,j,c}^{3s})} }}} } $$
\end{dmath}
where $z^3$ denotes the output of the fully connected layer $FC3$.


\par \textbf{Second object:} Unlike previous adversarial learning strategies \cite{Tzeng_2017_CVPR,pei2018multiadversarial}, the proposed method extends domain invariance to include intermediate representations between the domains. The optimization objective by considering RF-Mixup technique for different domains is specified as follows:

\begin{dmath}
    \label{lcdan}
    $${\mathcal{L}_{CDAN}} = \frac{1}{{{n_s}}}\sum\limits_{i = 1}^{{n_s}} {{w_{si}}\log \left( {D(f(x_i^s) \otimes g(f(x_i^s)))} \right)}  + \frac{1}{{{n_t}}}\sum\limits_{j = 1}^{{n_t}} {{w_{tj}}\log \left( {1 - D(f(x_j^t) \otimes g(f(x_j^t)))} \right)} + \frac{1}{{{N_b}}}\sum\limits_{k = 1}^{{N_b}} {\log (1 - D(e_k^{RFM} \otimes y_k^{RFM}))} $$
\end{dmath}

\par \textbf{Third object:} The third objective focuses on minimizing the distributional disparity between source and target domains. The PTPAI method utilizes the MK-MMSD distance criterion, implemented within two fully connected layers ($FC1$ and $FC2$), to match the domains, leveraging the properties of the kernel function within the RKHS. Based on Eq. \ref{MMSD3}, the MK-MMSD criterion is calculated using the following equations:

\begin{dmath}
    \label{dz1}
    $$\tilde d_{z1} = \frac{1}{{n_s^2}}\sum\limits_{i = 1}^{{n_s}} {\sum\limits_{j = 1}^{{n_s}} {{w_{si}}{w_{sj}}{K^2}(z_i^{1s},z_i^{1s})} }  + \frac{1}{{n_t^2}}\sum\limits_{i = 1}^{{n_t}} {\sum\limits_{j = 1}^{{n_t}} {{w_{ti}}{w_{tj}}{K^2}(z_i^{1t},z_i^{1t})} }  - \frac{2}{{{n_s}{n_t}}}\sum\limits_{i = 1}^{{n_s}} {\sum\limits_{j = 1}^{{n_t}} {{w_{si}}{w_{tj}}{K^2}(z_i^{1s},z_j^{1t})} } $$
\end{dmath}
\begin{dmath}
    \label{dz2}
    $$\tilde d_{z2} = \frac{1}{{n_s^2}}\sum\limits_{i = 1}^{{n_s}} {\sum\limits_{j = 1}^{{n_s}} {{w_{si}}{w_{sj}}{K^2}(z_i^{2s},z_i^{2s})} }  + \frac{1}{{n_t^2}}\sum\limits_{i = 1}^{{n_t}} {\sum\limits_{j = 1}^{{n_t}} {{w_{ti}}{w_{tj}}{K^2}(z_i^{2t},z_i^{2t})} }  - \frac{2}{{{n_s}{n_t}}}\sum\limits_{i = 1}^{{n_s}} {\sum\limits_{j = 1}^{{n_t}} {{w_{si}}{w_{tj}}{K^2}(z_i^{2s},z_j^{2t})} } $$
\end{dmath}
where $z^1$ and $z^2$ represent the outputs of layers $FC1$ and $FC2$, respectively. To effectively minimize the domain difference, multiple kernels with varying bandwidths are used to capture both low-order and high-order moments of the learned features. Eq. \ref{dz1} and Eq. \ref{dz2} utilize a linear combination of Gaussian kernels as follows:

\begin{dmath}
    \label{kernel}
    $$K \triangleq \sum\limits_{u = 1}^U {{\xi _u}{k_u}} $$
\end{dmath}
where $\xi$ represents the coefficients of each kernel, and $U$ represents the number of different kernels considered. The following equation is employed to compute the multi-layer MK-MMSD loss function $\mathcal{L}_{MK - MMSD}$:

\begin{dmath}
    \label{lmkmmsd}
    $${\mathcal{L}_{MK - MMSD}} = \tilde d_{z1} + \tilde d_{z2} $$
\end{dmath}

\par \textbf{Fourth object:} The fourth objective aims to minimize the losses associated with the auxiliary classifier and discriminator. Eq. \ref{aux-class} and Eq. \ref{aux-dis} are employed  accomplish this purpose. The auxiliary objective function ($\mathcal{L}_{Aux}$) is stated as follows:

\begin{dmath}
    \label{laux}
    $${\mathcal{L}_{Aux}} = {\mathcal{L}_{{{\tilde G}_y}}} + {\mathcal{L}_{{{\tilde G}_d}}}$$
\end{dmath}

\par Given all objective functions, the PTPAI method's ultimate optimization objective is specified as:

\begin{dmath}
    \label{total}
    $${\mathcal{L}_{total}}\left( {\Theta _f^*,\Theta _c^*,\Theta _d^*,\Theta _a^*} \right) = {\mathcal{L}_C}\left( {{\Theta _f},{\Theta _c}} \right) - \mu {\mathcal{L}_{CDAN}}\left( {{\Theta _f},{\Theta _d}} \right) + \gamma {\mathcal{L}_{MK - MMSD}}\left( {{\Theta _f}} \right) + {\mathcal{L}_{Aux}}({\Theta _a})$$
\end{dmath}
where $\mu$ and $\gamma$ are hyperparameters controlling the trade-off between the four terms. ${\Theta _f}$, ${\Theta _c}$, ${\Theta _d}$, and ${\Theta _a}$ denote the parameters of the feature extractor, label classifier, domain discriminator, and auxiliary domain discriminator respectively. ${\Theta _f^*}$, ${\Theta _c^*}$, ${\Theta _d^*}$, and ${\Theta _a^*}$ represent the optimal values of these parameters. The PTPAI method parameters are updated by the Adam optimizer and back-propagation techniques in the following ways:

\begin{dmath}
    \label{tetaF}
    $${\Theta _f} \leftarrow {\Theta _f} - \eta \left( {\frac{{\partial {\mathcal{L}_C}}}{{\partial {\Theta _f}}} - \mu \frac{{\partial {\mathcal{L}_{CDAN}}}}{{\partial {\Theta _f}}} + \gamma \frac{{\partial {\mathcal{L}_{MK - MMSD}}}}{{\partial {\Theta _f}}}} \right)$$
\end{dmath}

\begin{dmath}
    \label{tetaC}
    $${\Theta _c} \leftarrow {\Theta _c} - \eta \left( {\frac{{\partial {\mathcal{L}_C}}}{{\partial {\Theta _c}}}} \right)$$
\end{dmath}

\begin{dmath}
    \label{tetaD}
    $${\Theta _d} \leftarrow {\Theta _d} - \eta \left( {\frac{{\partial {\mathcal{L}_{CDAN}}}}{{\partial {\Theta _d}}}} \right)$$
\end{dmath}

\begin{dmath}
    \label{tetaA}
    $${\Theta _a} \leftarrow {\Theta _a} - \eta \left( {\frac{{\partial {\mathcal{L}_{Aux}}}}{{\partial {\Theta _a}}}} \right)$$
\end{dmath}
where $\eta$ represents the learning rate. Algorithm 1 is a brief overview of the pseudo-code for the proposed PTPAI method.

 \begin{figure}[t]

    \centering
    \includegraphics[width=\columnwidth]{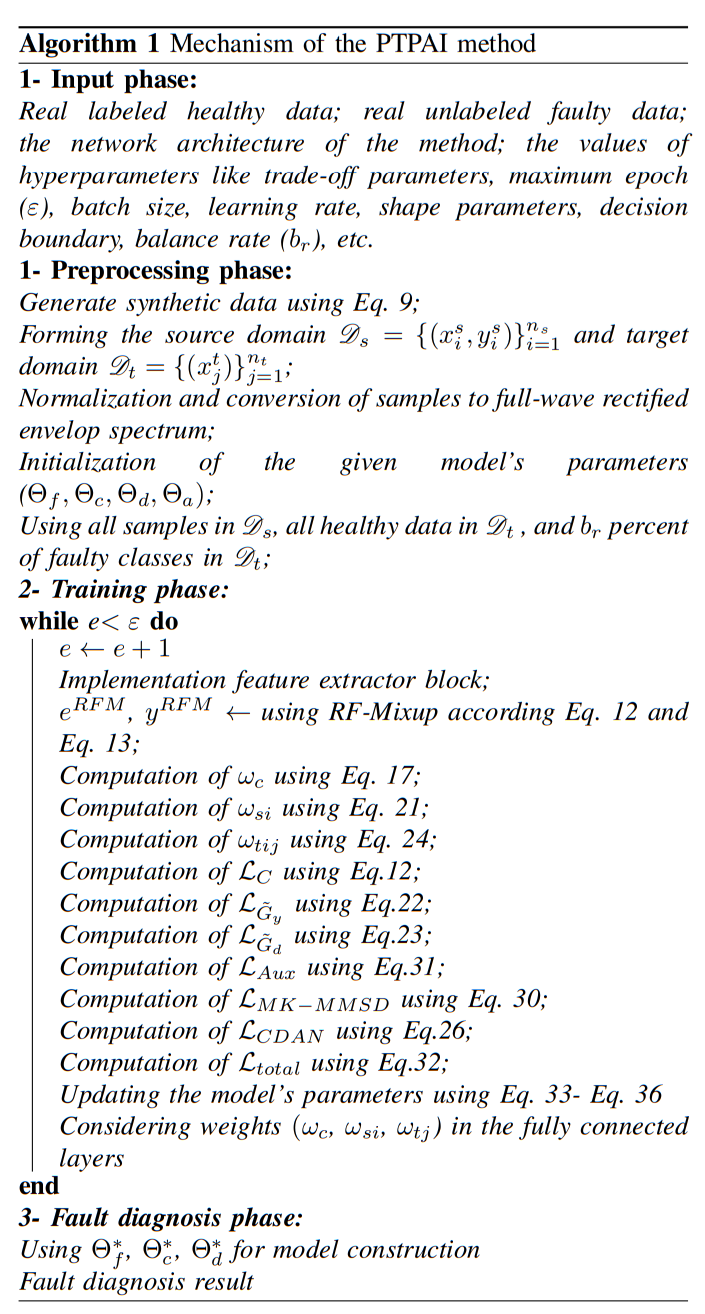}
    \label{algorithm1}
    \end{figure}

\section{ Experimental studies and comparison}

\subsection{Dataset description}
This section investigates the efficiency of the suggested method for diagnosing bearing faults in the presence of missing data and employing synthetic data. To assess the effectiveness and validity of the proposed PTPAI method, two widely recognized datasets, CWRU \cite{LOU20041077} and JNU \cite{li2019novel}, are employed for comprehensive evaluation.

\subsubsection{CWRU}
The CWRU dataset comprises vibration signals acquired from the drive-end accelerometer of a test rig, sampled at a frequency of 12 kHz and is divided into four health states: normal condition (NC), inner race fault (IRF), ball fault (BF), and outer race fault (ORF). To ensure consistency, sub-fault types of varying fault severities are grouped, resulting in 4800 samples for the actual dataset, with each health state comprising 1200 sampled segments of 4096 points. To generate realistic faults, we only use actual healthy data. Half of the healthy samples are designated as synthetic source data to prevent data leakage during evaluation. The number of samples in the healthy class for both domains is then up-sampled to 1200.

\subsubsection{JNU}
The JNU dataset comprises vibration data and focuses on a single-row spherical roller bearing and includes intentionally induced faults using wire-cutting machines. The data collection process involved four health conditions: NC, IRF, ORF, and BF. The data was acquired over a 30-second interval at a sampling rate of 50 kHz. This comprehensive dataset provides diverse fault scenarios, enabling in-depth analysis and evaluation in the present investigation. The healthy class in this dataset maintains the same sample size as the previous dataset. Detailed information regarding the transfer tasks is presented in Table \ref{CWRU_DATA}.

\begin{table}[]
\caption{Description of transfer tasks considering missing data in CWRU and JNU datasets. Tasks C  and J are related to CWRU and JNU datasets, respectively.}
\label{CWRU_DATA}
\resizebox{\columnwidth}{!}{%
\begin{tabular}{cccc}
\hline
Task & source domain         & target domain & balance rate                         \\ \hline
C1 or J1   & All synthetic classes & All real data & Completely balanced, 10\% , 5\%, 1\% \\
C2 or J2   & All synthetic classes & NC,IRF,BF     & Completely balanced, 10\% , 5\%, 1\% \\
C3 or J3   & All synthetic classes & NC,BF,ORF     & Completely balanced, 10\% , 5\%, 1\% \\
C4 or J4   & All synthetic classes & NC,IRF        & Completely balanced, 10\% , 5\%, 1\% \\
C5 or J5   & All synthetic classes & NC,ORF        & Completely balanced, 10\% , 5\%, 1\% \\
C6 or J6   & All synthetic classes & IRF           & Completely balanced, 10\% , 5\%, 1\% \\
C7 or J7   & All synthetic classes & BF            & Completely balanced, 10\% , 5\%, 1\% \\ \hline
\end{tabular}%
}
\end{table}

\subsection{Comparison method}
To evaluate the proposed PTPAI method, it is compared to a variety of SOTA methods, including deep transfer learning based convolutional neural network (DTLCNN) \cite{8809786}, domain adversarial neural network (DANN) \cite{ajakan2015domainadversarial}, CDAN \cite{long2018conditional}, simulation-driven subdomain adaptation network (SDSAN) \cite{liu2023simulation}, class-weighted alignment based transfer network (CWATN) \cite{zhang2022partial}, Physics-informed Artificial Intelligence with Partial term (PAIP), Physics-informed Artificial Intelligence with Regularization term (PAIR), 
Physics-informed Artificial Intelligence with Maximum mean discrepancy term (PAIM), Mixup, Manifold Mixup, and CutMix. For a fair comparison, the baseline of each of these methods is similar to the proposed method. These techniques fall into the following categories:

\begin{enumerate}
\item \textbf{Domain adaptation (DA)-based methods}: DTLCNN, DANN, and CDAN methods are representative of this category. DTLCNN utilizes multi-layer multi-kernel MMD in its fully connected layers. DANN employs an adversarial domain discriminator to lessen the distribution sicrepancy between domains. CDAN aligns the distribution of domains using label information. However, this category of techniques lacks a dedicated module to address the missing data problem.
\item \textbf{Subdomain adaptation (SDA-based methods}: This group includes the SDSAN technique. The SDSAN method incorporates the LMMD criterion as a subdomain adaptation module to matcj subdomains with identical classes across domains. This category of methods lacks a module to address the detrimental effects of outlier classes and imbalanced datasets.
\item \textbf{Partial transfer learning (PTL)-based methods}: CWATN and PAIP methods try to decrease the effect of outlier classes in the target domain by considering weight coefficients. The PAIP method has the identical structural design as the PTPAI method, excluding using the RF-MIXUP technique. The CWATN technique adjusts the global distribution of domains using the MK-MMD module and employs weighted class-level aligning to lessen the impact of outlier classes.
\item \textbf{Regularization-based methods}: The PAIR method falls within this category. This method is similar to the proposed method, but there is no trace of the Weighting block. Also, to facilitate a more comprehensive evaluation of the RF-MIXUP regularization technique, this method will be compared with some state-of-the-art regularization approaches, including Mixup, Manifold Mixup, and CutMix.
\item \textbf{Hybrid partial and regularization (HPR)-based method}: This category includes the PAIM approach. With one exception, this technique is similar to our suggested approach. Instead of MK-MMSD, the PAIM technique uses MK-MMD. By comparing the suggested approach to the PAIM method, the effect of utilizing MK-MMSD can be evaluated.
\end{enumerate}

\subsection{Implementation details}

The research conducted in this study involves the evaluation of various techniques using an identical amount of missing data in the target domain. The training phase for all methods consists of 200 epochs, with a batch size of 128. The learning rate is dynamically adjusted throughout the learning process and follows a decreasing function defined by the equation $\eta = {{l{r_0}} \mathord{\left/{\vphantom {{l{r_0}} {{{\left( {1 + 10i} \right)}^{0.75}}}}} \right.\kern-\nulldelimiterspace} {{{\left( {1 + 10i} \right)}^{0.75}}}}$, where $lr_0$ is the initial learning rate, and $i$ is the percentage of iterations. A smaller value for $lr_0$ is generally preferred due to the influence of the stochastic nature of the $\beta$ distribution on the generated features. Consequently, a value of 0.002 is selected for $lr_0$ in the proposed method. To enhance experimentation and analysis, penalty coefficients $\mu$ and $\gamma$ are assumed to take values from the set $\{0, 0.02, 0.05, 0.1, 0.5, 1\}$. The optimal values of $\mu=1$ and $\gamma=0.05$ are determined through experimental investigations. In the training phase, the network's parameters are initialized using the Xavier initializer \cite{KavianEARTH2021}, and the Adam optimizer \cite{MultiScaleKavianpour} is employed for the training process. The datasets are partitioned into training, validation, and test sets, with proportions of 80\%, 10\%, and 10\%, respectively.  Moreover, the proposed PTPAI approach utilizes Multi-Gaussian kernels with five distinct bandwidths $\{0.001, 0.01, 1, 10, 100\}$ for the MK-MMSD algorithm. All experiments are conducted using the PyTorch framework, and to account for random variations, each experiment is replicated ten times. Additionally, all hyperparameters are determined through a grid search technique. To achieve a fair comparison, the same structure and training conditions settings are employed across all approaches.

\subsection{Evaluation metrics}
To assess the effectiveness of the suggested methodology, two metrics, namely balanced-accuracy (b-accuracy) and F1-score, have been employed \cite{CHAMLAL2024111367}. In the context of imbalanced datasets, it is well acknowledged that standard accuracy measures may not offer a full assessment of the classification performance for individual classes.
As a result, we employ a balanced version of accuracy. The b-accuracy criterion is computed as a average of specificity and sensitivity in the following manner:
\begin{equation}
    Sensitivity = \frac{TP}{TP + FN}
\end{equation}
\begin{equation}
    Specificity = \frac{TN}{TN + FP}
\end{equation}

\begin{equation}
    b-accuracy = \frac{1}{2}\left( {Sensitivity + Specificity} \right)
\end{equation}

The variables TP, TN, FP, and FN indicate the quantities of true positive, true negative, false positive, and false negative, respectively.
The F1-score is a metric commonly used to evaluate the problem of imbalanced datasets. It is derived based on the following relationship:
\begin{equation}
    \begin{aligned}
       F1 = \frac{{2TP}}{{2TP + FP + FN}}
    \end{aligned}
\end{equation}
The metrics employed in this study offer a reliable means of comparing imbalanced datasets, and the given results are derived from the average of 10 repeats.

\begin{figure*}[t]
\begin{subfigure}{.5\textwidth}
  \centering
  \includegraphics[width=\columnwidth]{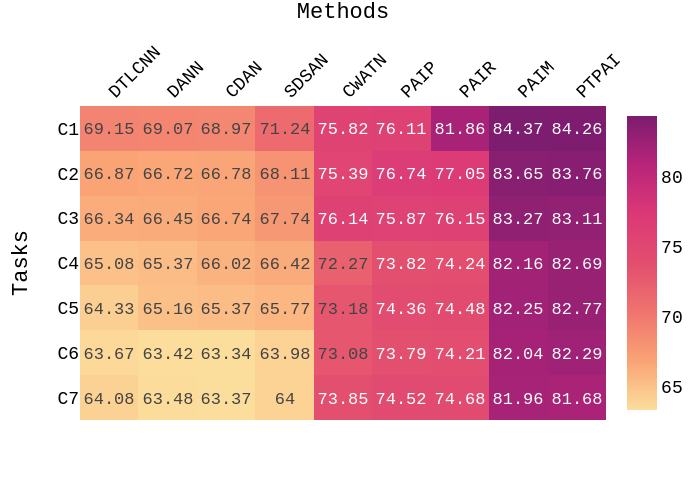}  
  \caption{The b-accuracy of fault diagnosis for a balancing rate of 1\%}
  \label{fig:sub-first}
\end{subfigure}
\begin{subfigure}{.5\textwidth}
  \centering
  \includegraphics[width=\columnwidth]{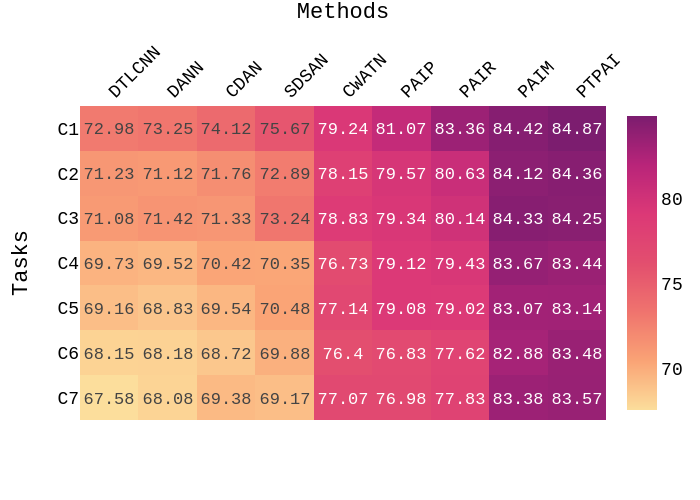}
  \caption{The b-accuracy of fault diagnosis for a balancing rate of 5\%}
  \label{fig:sub-second}
\end{subfigure}

\begin{subfigure}{.5\textwidth}
  \centering
  \includegraphics[width=\columnwidth]{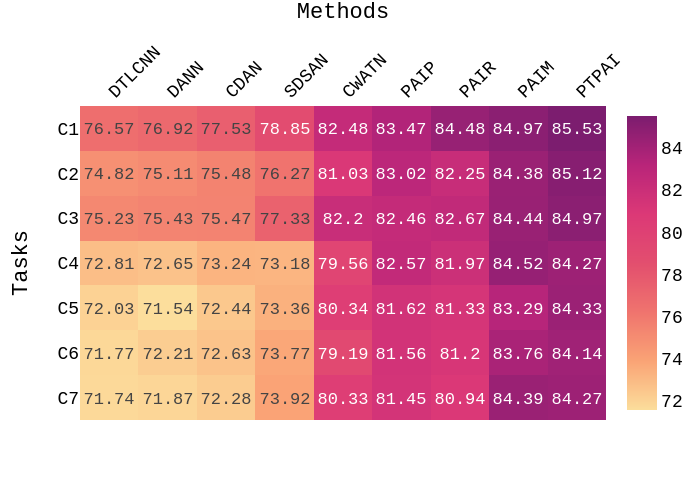}  
  \caption{The b-accuracy of fault diagnosis for a balancing rate of 10\%}
  \label{fig:sub-third}
\end{subfigure}
\begin{subfigure}{.5\textwidth}
  \centering
  \includegraphics[width=\columnwidth]{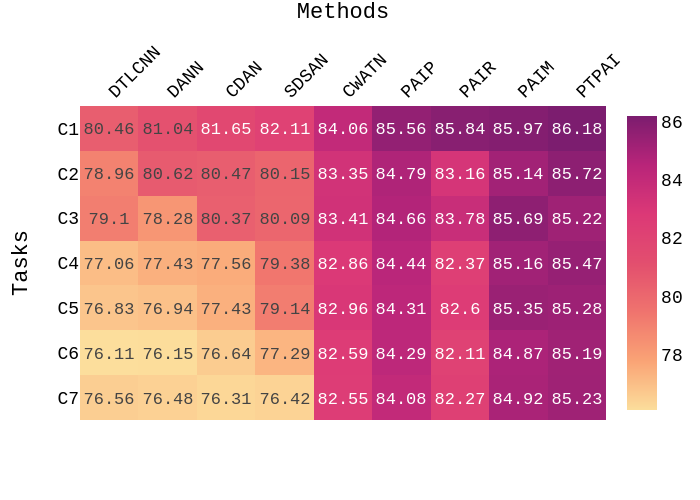}  
  \caption{The b-accuracy of fault diagnosis for a fully balanced condition}
  \label{fig:sub-fourth}
\end{subfigure}
\caption{Different techniques' diagnosis b-accuracy for different ranges of balance rate in the CWRU dataset}
\label{c_acc}
\end{figure*}

\begin{figure*}[t]
\begin{subfigure}{.5\textwidth}
  \centering
  \includegraphics[width=\columnwidth]{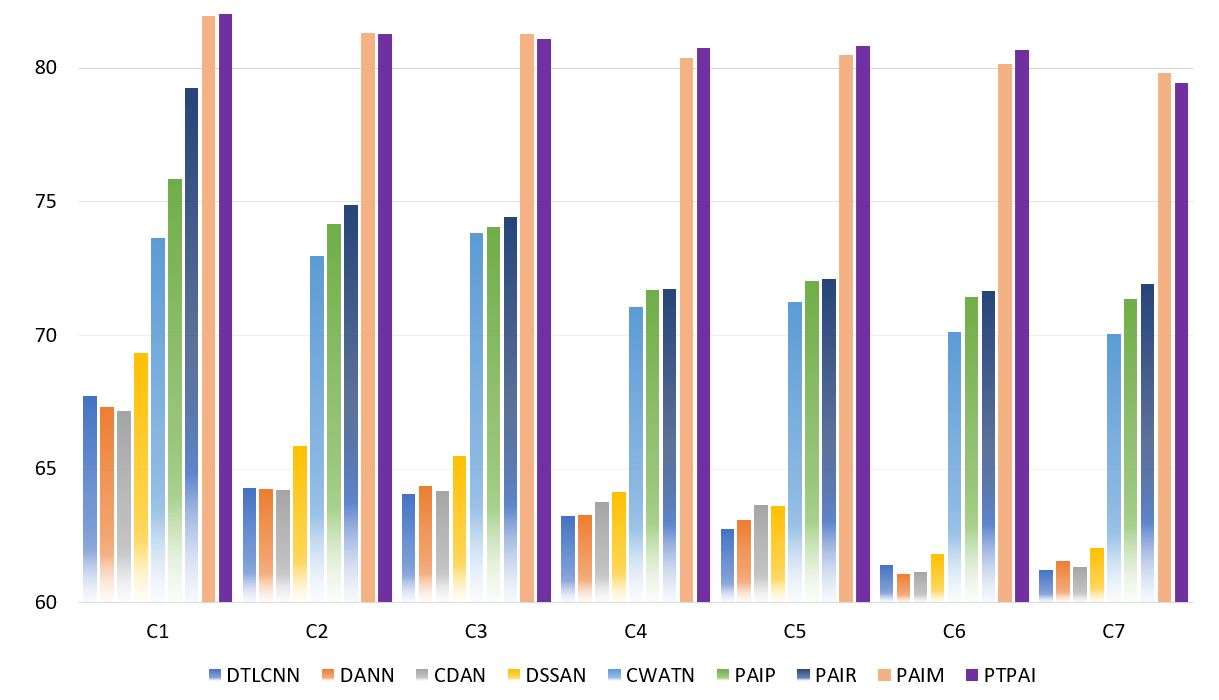}  
  \caption{The F1-score of fault diagnosis for a balancing rate of 1\%}
  \label{fig:sub-first}
\end{subfigure}
\begin{subfigure}{.5\textwidth}
  \centering
  \includegraphics[width=\columnwidth]{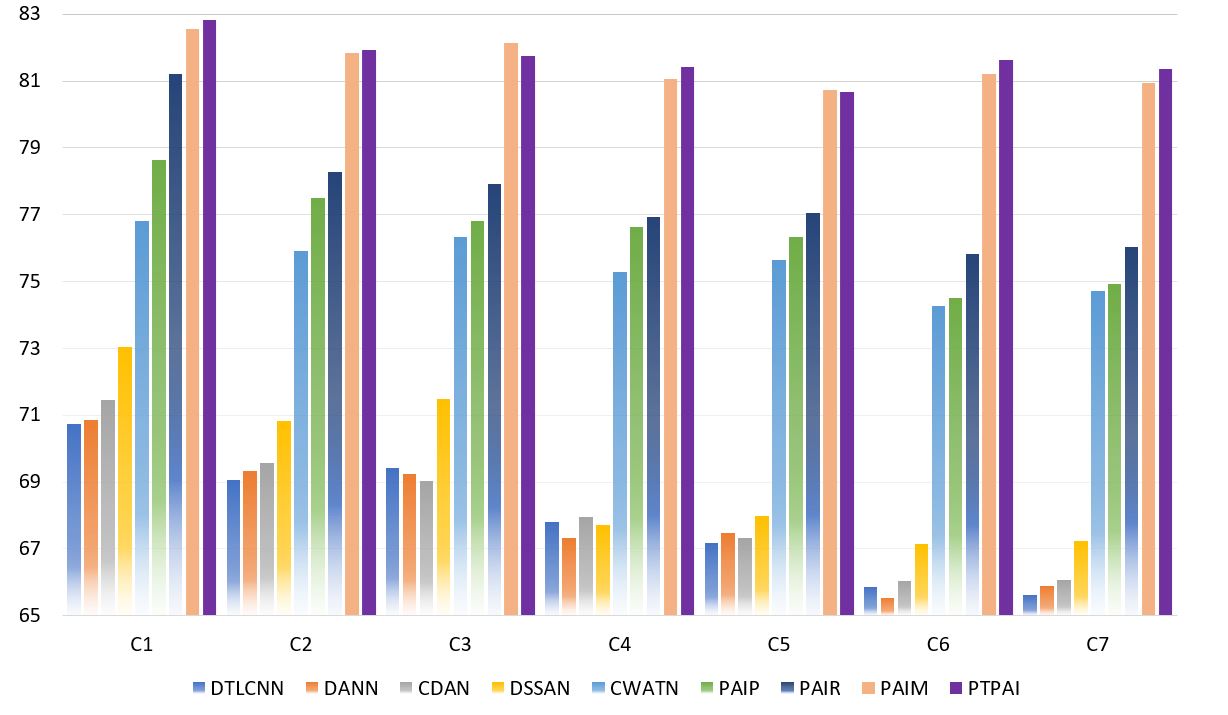}
  \caption{The F1-score of fault diagnosis for a balancing rate of 5\%}
  
\end{subfigure}

\begin{subfigure}{.5\textwidth}
  \centering
  \includegraphics[width=\columnwidth]{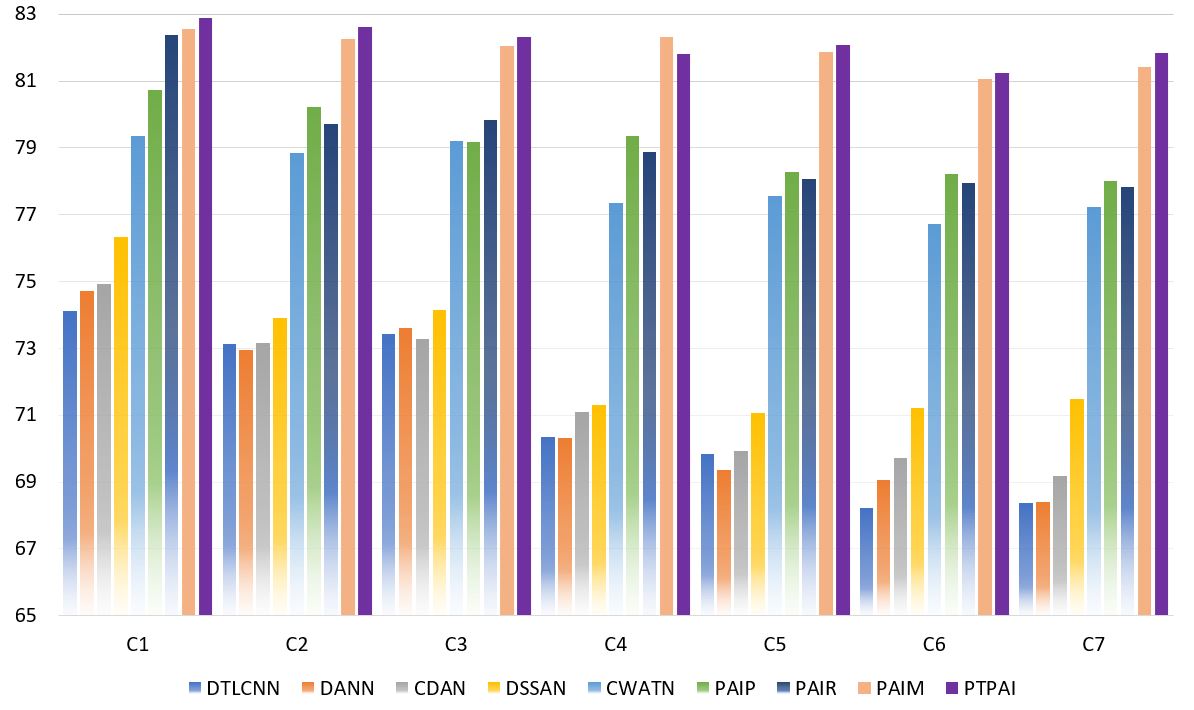}  
  \caption{The F1-score of fault diagnosis for a balancing rate of 10\%}
  \label{fig:sub-third}
\end{subfigure}
\begin{subfigure}{.5\textwidth}
  \centering
  \includegraphics[width=\columnwidth]{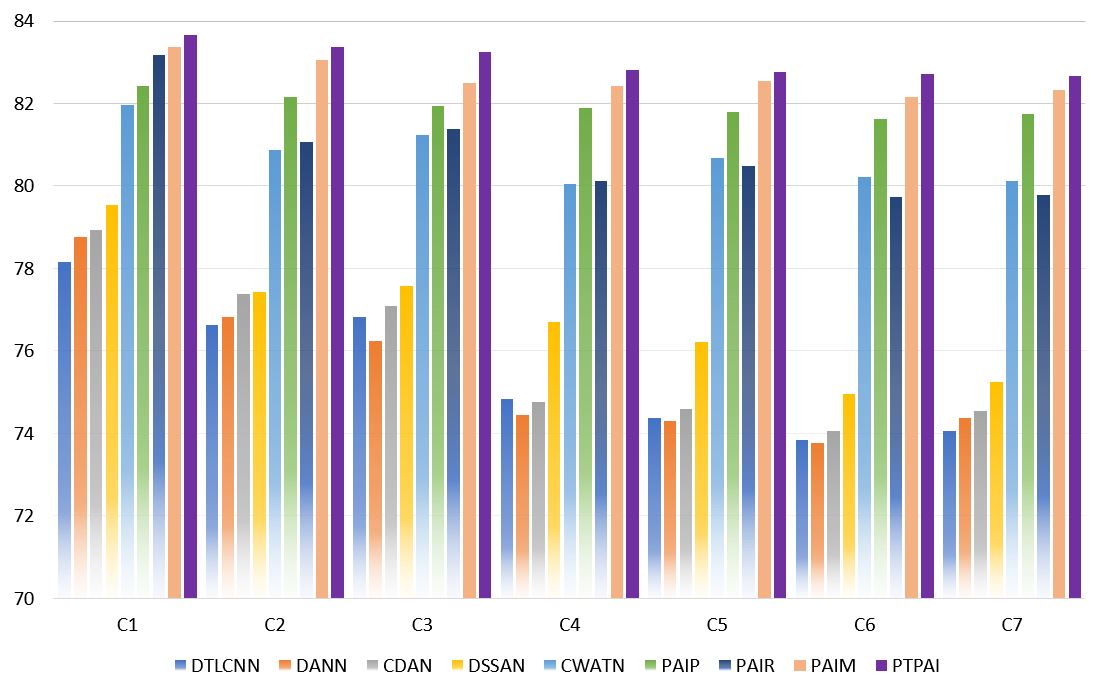}  
  \caption{The F1-score of fault diagnosis for a fully balanced condition}
  \label{fig:sub-fourth}
\end{subfigure}
\caption{Different techniques' diagnosis F1-score for different ranges of balance rate in the CWRU dataset}
\label{c-f1}
\end{figure*}

\section{Result}
This section assesses the efficacy of the PTPAI technique on the CWRU and JNU datasets in the presence of missing data. In this case, it is assumed that the amount of imbalance in all defective classes in the target domain is the same. For instance, in the CWRU dataset, task $C1$ with a 1\% balance level in the target domain contains 1200 NC class samples (100\%), 12 IRF class samples (1\%), 12 ORF class samples (1\%), and 12 BF class samples (1\%). The lower the value of the level of balance, the number of samples of faulty classes will be less than that of the healthy class. In addition to different balance levels, quantitative analysis under different missing class settings has been done.
Across all experimental modes, access to 100\% of NC class samples is assumed, a valid hypothesis in real-world applications. The target domain lacks outlier classes, while the source domain exhibits an imbalanced class distribution.

\begin{figure*}[t]
\begin{subfigure}{.5\textwidth}
  \centering
  \includegraphics[width=\columnwidth]{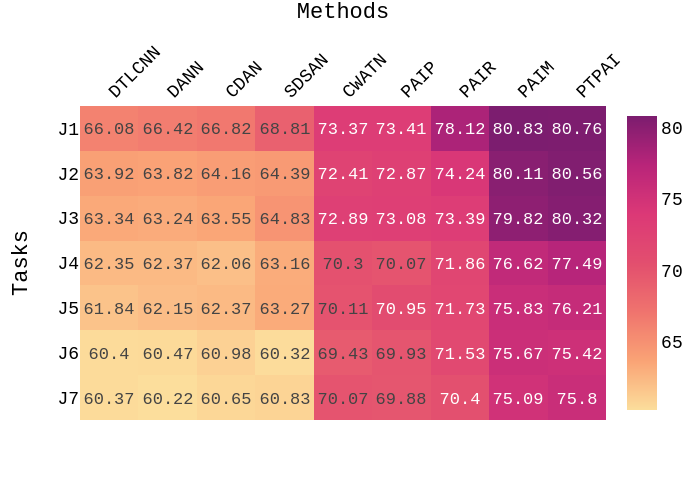}  
  \caption{The b-accuracy of fault diagnosis for a balancing rate of 1\%}
  \label{fig:sub-first}
\end{subfigure}
\begin{subfigure}{.5\textwidth}
  \centering
  \includegraphics[width=\columnwidth]{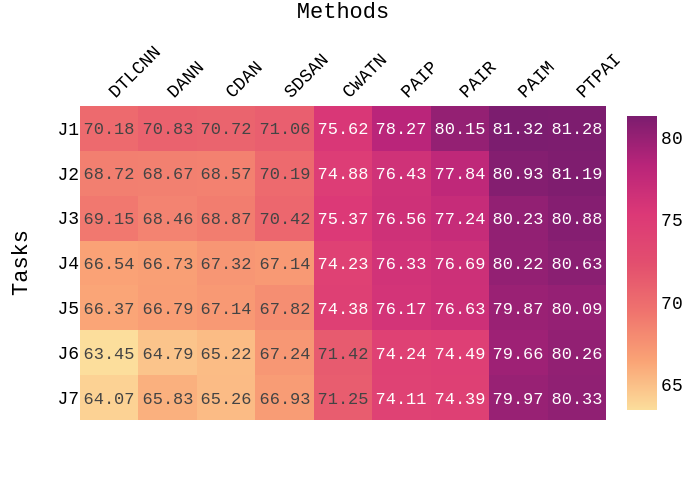}
  \caption{The b-accuracy of fault diagnosis for a balancing rate of 5\%}
  \label{fig:sub-second}
\end{subfigure}

\begin{subfigure}{.5\textwidth}
  \centering
  \includegraphics[width=\columnwidth]{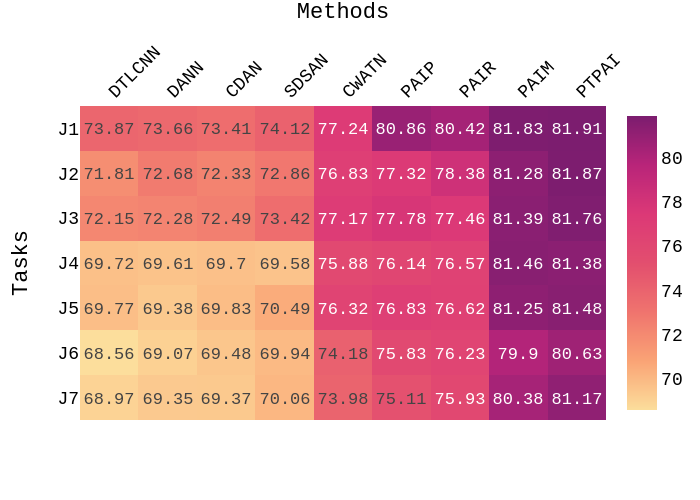}  
  \caption{The b-accuracy of fault diagnosis for a balancing rate of 10\%}
  \label{fig:sub-third}
\end{subfigure}
\begin{subfigure}{.5\textwidth}
  \centering
  \includegraphics[width=\columnwidth]{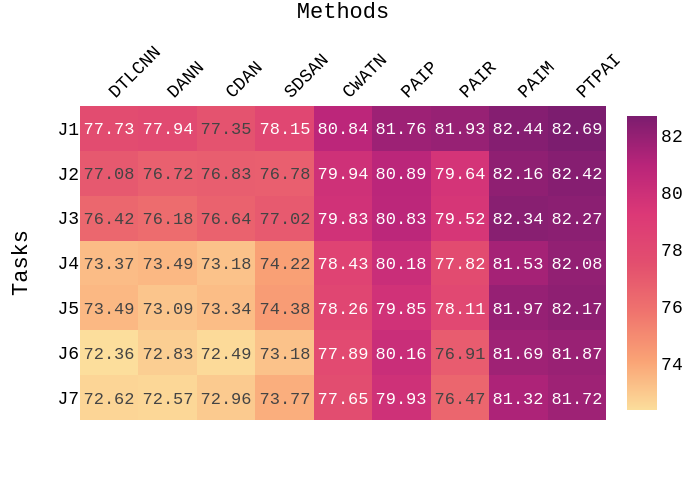}  
  \caption{The b-accuracy of fault diagnosis for a fully balanced condition}
  \label{fig:sub-fourth}
\end{subfigure}
\caption{Different techniques' diagnosis b-accuracy for different ranges of balance rate in the JNU dataset}
\label{j-acc}
\end{figure*}

\begin{figure*}[t]
\begin{subfigure}{.5\textwidth}
  \centering
  \includegraphics[width=\columnwidth]{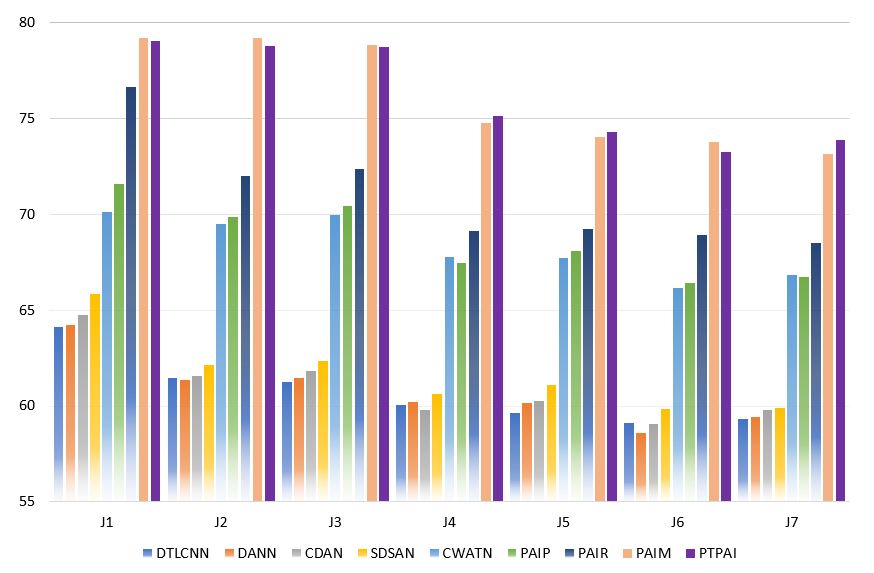}  
  \caption{The F1-score of fault diagnosis for a balancing rate of 1\%}
  \label{fig:sub-first}
\end{subfigure}
\begin{subfigure}{.5\textwidth}
  \centering
  \includegraphics[width=\columnwidth]{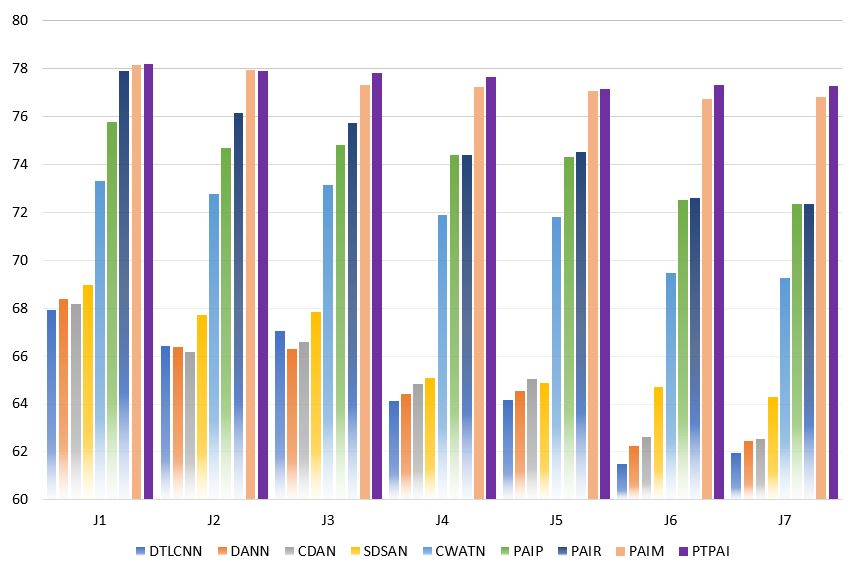}
  \caption{The F1-score of fault diagnosis for a balancing rate of 5\%}
  \label{fig:sub-second}
\end{subfigure}

\begin{subfigure}{.5\textwidth}
  \centering
  \includegraphics[width=\columnwidth]{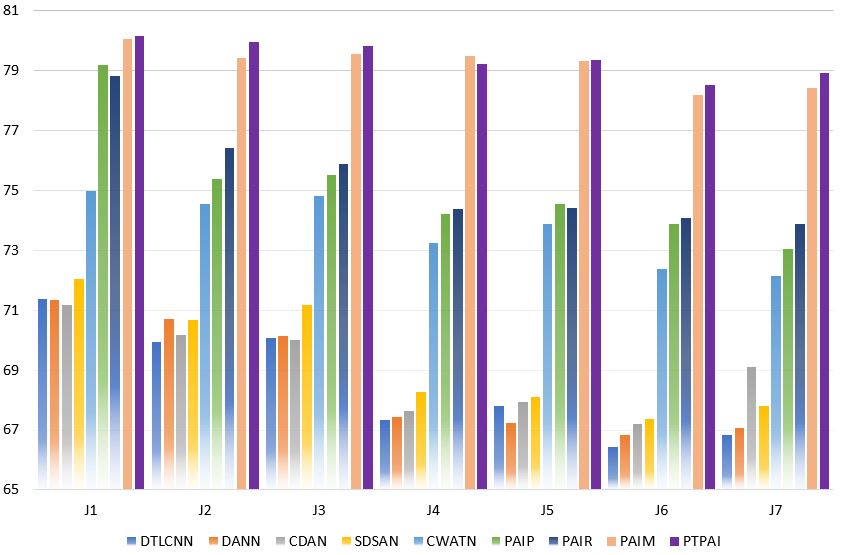}  
  \caption{The F1-score of fault diagnosis for a balancing rate of 10\%}
  \label{fig:sub-third}
\end{subfigure}
\begin{subfigure}{.5\textwidth}
  \centering
  \includegraphics[width=\columnwidth]{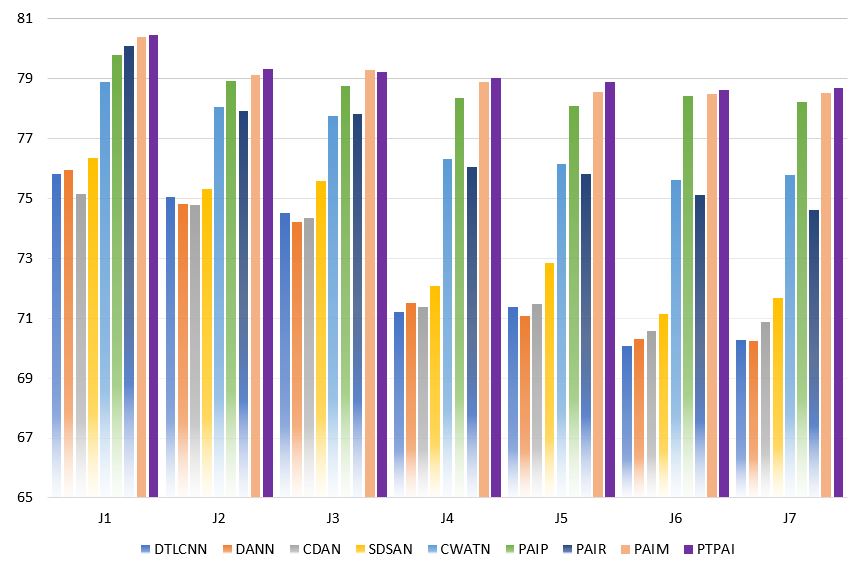}  
  \caption{The F1-score of fault diagnosis for a fully balanced condition}
  \label{fig:sub-fourth}
\end{subfigure}
\caption{Different techniques' diagnosis F1-score for different ranges of balance rate in the JNU dataset}
\label{j-f1}
\end{figure*}

\subsection{Case study 1: CWRU dataset}

\par The experimental results for bearing fault diagnosis, considering both the synthetic dataset and the real dataset of CWRU, are presented in Figure \ref{c_acc} based on the b-accuracy criterion and in Figure \ref{c-f1} based on the F1-score criterion. Notably, the PTPAI method consistently suppresses comparative methods across the majority of tasks with regarding different balance levels and the presence of outlier classes in the source domain. This observation is highly encouraging and reinforces the effectiveness of the PTPAI approach. The collected results can be categorized into the following groups:
\begin{enumerate}
    \item In the majority of tasks, the proposed method and HPR-based method have better performance than the PAIM method. This improvement is attributed to the utilization of MK-MMSD instead of MK-MMD, enabling better capture of both variance and average data information in the RKHS. Consequently, the suggested approach successfully lessens the disparity in distribution between domains.
    \item In the context of the CDFD problem (task C1), a significant decrease in performance is observed for DA-based, SDA-based, and PTL-based methods as the balance rate decreases from full balance to 1\%. For example, the F1-score of the SDSAN method decreases by 10.21\% in this scenario. Similarly, the PAIP and PTPAI methods experience reductions of 6.59\% and 1.65\%, respectively. The proposed method demonstrates good robustness to changes in the balance level, highlighting the effectiveness of the RF-MIXUP technique for addressing imbalanced datasets. Additionally, the proposed method consistently performs well in PSFD tasks across different balance levels.
    \item In the balanced mode and within the context of the closed-set problem denoted as $C1$, the proposed method achieves a b-accuracy of 86.18\%. This result is 0.34\% higher compared to the PAIR method, demonstrating the best performance among the comparison methods in this specific condition. Furthermore, the F1-score of the proposed method is 83.67\%, surpassing the PAIR method by 0.50\% in terms of performance. The relatively low values of b-accuracy and F1-score in this completely balanced task indicate a significant distribution disparity between the real and synthetic data. Additionally, the results of this task demonstrate the proposed method's superior performance compared to conventional DA-based and SDA-based methods in physics-informed AI problems.
    \item In the completely balanced condition, the presence of outlier classes in the source domain significantly reduces the performance of conventional DA-based and SDA-based methods in PSFD problems. However, the inclusion of a weighting mechanism effectively mitigates the detrimental impact of outlier class samples in the source domain, enabling the focus on the shared class space within the dataset during domain or subdomain adaptation. The utilization of instance-level and class-level weighting blocks contributes to the superior performance of the PAIP method compared to CWATN in this scenario. Additionally, the incorporation of a regularization block, in conjunction with an appropriate weighting block, further enhances the overall performance of the PTPAI and PAIM methods, proving advantageous over PTL and regularization-based methods.
    \item The PTPAI approach surpasses other methods in the majority of tasks, as evidenced by higher b-accuracy and F1-score metrics. This demonstrates the superior performance of the PTPAI approach across diverse balance levels and in the presence of outlier classes. Conversely, conventional methods experience a significant decline in performance as the number of outlier classes and the severity of imbalance increase. The proposed PTPAI method exhibits good robustness and maintains its performance even under challenging conditions with increased outlier classes and imbalance.
\end{enumerate}

\subsection{Case study 2: JNU dataset}

Figures \ref{j-acc} and \ref{j-f1}  indicate the experimental outcomes for the JNU dataset based on the b-accuracy and F1-score metrics, respectively. Similar to the findings for the CWRU dataset, the recommended strategy outperforms comparative strategies. The outcomes can be classified into the following groups:
\begin{enumerate}
    \item As the balance level decreases, the comparative methods' performance for the first three categories exhibits a more pronounced decline compared to the rest of methods. For instance, in the $C7$ task, when reducing the balance level from 5\% to 1\%, the SDSAN method's b-accuracy drops from 69.67\% to 64.36\%. In contrast, the proposed method maintains a higher level of b-accuracy, decreasing from 83.57\% to 83.12\%. This highlights the robustness of the PTPAI approach in handling imbalanced datasets. The improved performance of the proposed method can be attributed to the incorporation of the RF-Mixup technique as a regularization method.
    \item As the number of outlier classes in the source domain increases, the accuracy of the comparative methods decreases more significantly compared to the proposed method. For instance, in the balanced case and for the b-accuracy metric, the b-accuracy of the PTPAI method decreased by 4.96\% from task J1 to J7, while the b-accuracies of the PTL-based, regularization-based, SDA-based, and DA-based methods decreased by 1.83\%, 5.46\%, 4.97\%, and 4.86\%, respectively. This demonstrates the robustness of the PTPAI method to the increase in outlier classes in the source domain, which can be attributed to the presence of the weighting block in this method.
    \item In the case where there is no missing data, i.e., task J1 for the completely balanced condition, the PTPAI method suppresses other methods. the F1-score of the PTPAI method surpasses that of the SDSAN method, a classic SDA-based approach, by 4.57\%. This indicates the appropriate performance of the PTPAI method in scenarios involving synthetic data.
\end{enumerate}

\subsection{Discussion and analysis}
\subsubsection{Comparison different regularization techniques}
A comparative ablation study was conducted to assess the effectiveness of RF-Mixup compared to other well-known regularization techniques, including CutMix, Manifold Mixup, and Mixup. The evaluation was conducted based on the F1-score metric, which measures the diagnostic performance. The results for tasks $C6$ and $J3$ with balance rates of 1\% and 10\% are presented in Table \ref{reg}. These results clearly demonstrate the superiority of the RF-Mixup technique over the other approaches. RF-Mixup proposes a decoupling of the mixing factors utilized in Mixup, Manifold Mixup, and CutMix. It incorporates a weighting scheme that assigns higher weights to the minority class, thus favoring the minority class during the labeling process. This weighting mechanism proves beneficial in imbalanced circumstances. The method includes criteria that regulate the extent to which the minority features influence the label of the generated features, allowing for a trade-off between the majority and minority classes. As a result, RF-Mixup consistently achieves better performance across various balance levels in different tasks.

\begin{table}[h]
\caption{The F1-score comparison of the proposed method with other regularization methods}
\label{reg}
\resizebox{\columnwidth}{!}{%
\begin{tabular}{lcccc}
\hline
\multicolumn{1}{c}{Tasks} & Mixup & Manifold Mixup & CutMix & RF-Mixup \\ \hline
C6 (1\%)                  & 78.32 & 79.07          & 80.33   & 80.67    \\
C6 (10\%)                 & 79.88 & 80.34          & 81.11  & 81.23    \\
J3 (1\%)                  & 76.33 & 77.39          & 78.58  & 78.76    \\
J3 (10\%)                 & 78.52 & 79.05          & 79.77  & 79.82    \\ \hline
\end{tabular}%
}
\end{table}

\subsubsection{Analysis of weighting block}

\begin{figure}[th]
\centering

\begin{subfigure}{.5\columnwidth}
  \centering
  \includegraphics[width=\linewidth]{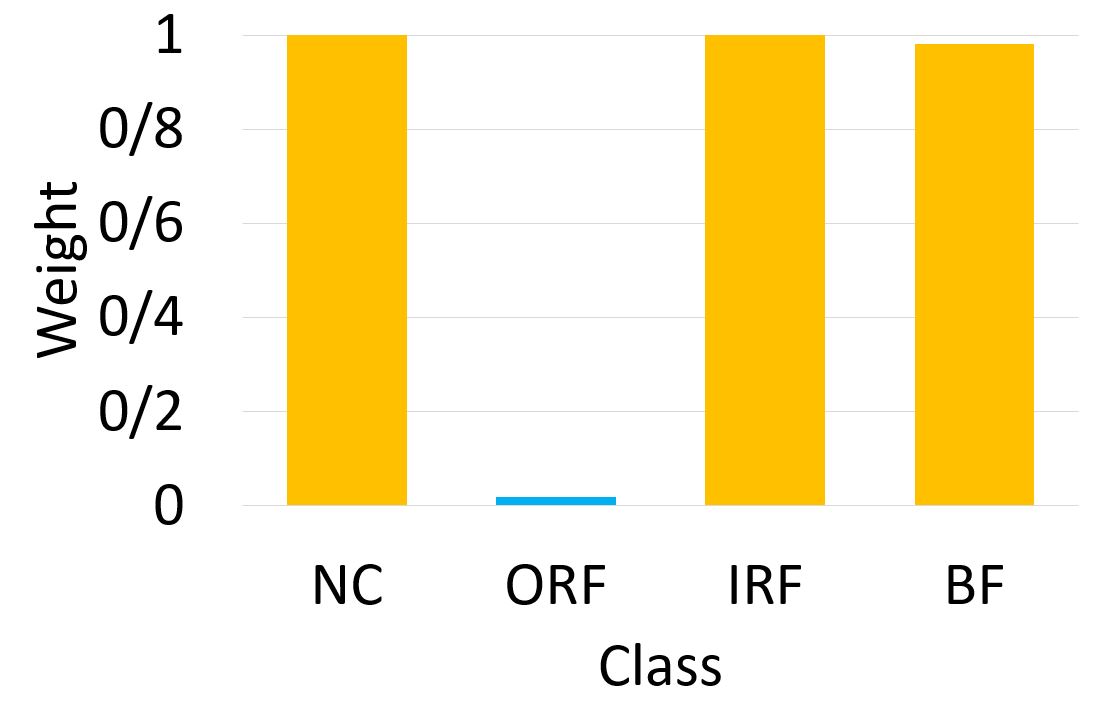} 
  \caption{}
  \label{weighting-a}
\end{subfigure}%
\begin{subfigure}{.5\columnwidth}
  \centering
  \includegraphics[width=\linewidth]{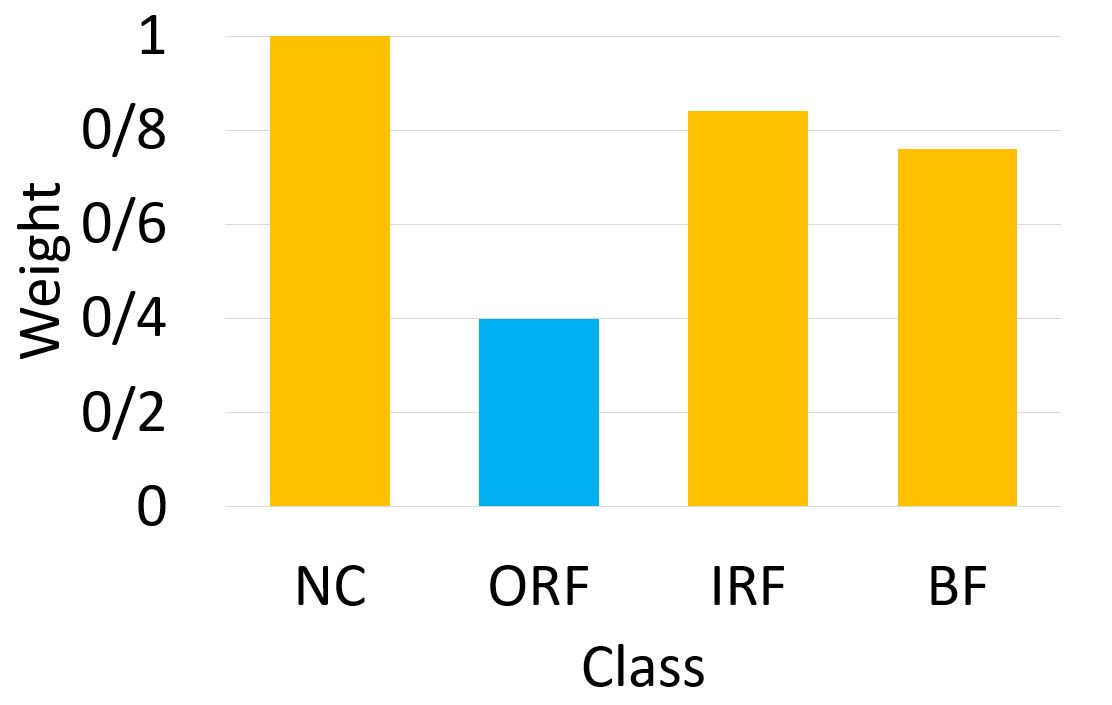}
  \caption{}
  \label{weighting-b}
\end{subfigure}

\begin{subfigure}{.5\columnwidth}
  \centering
  \includegraphics[width=\linewidth]{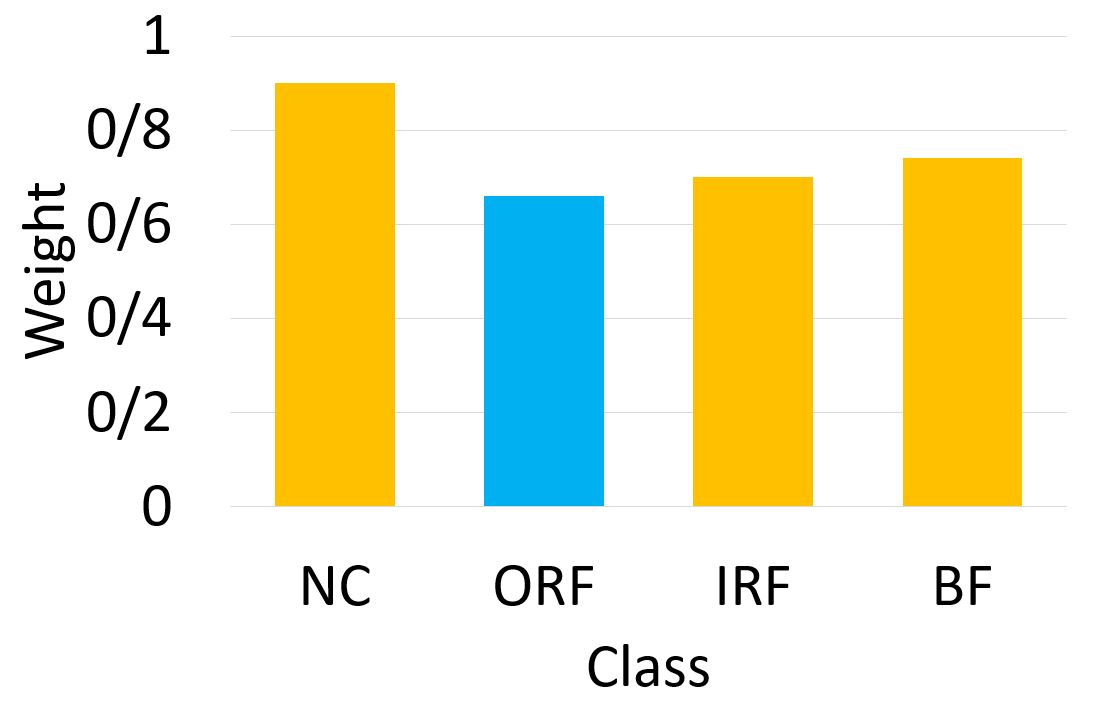}
  \caption{}
  \label{weighting-c}
\end{subfigure}%
\begin{subfigure}{.5\columnwidth}
  \centering
  \includegraphics[width=\linewidth]{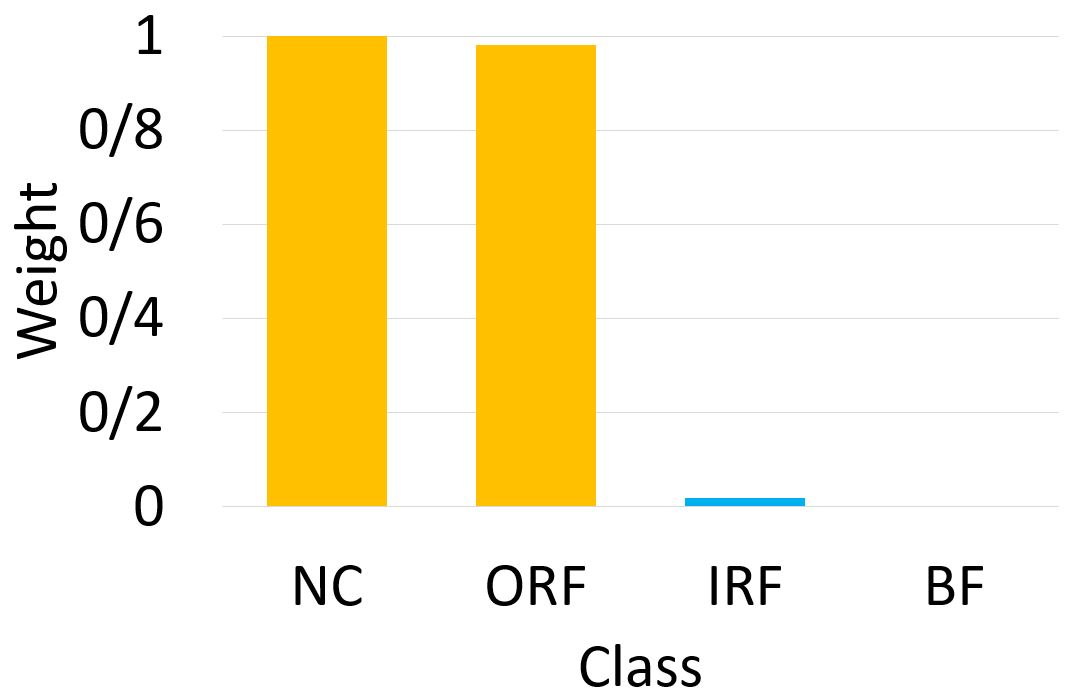}
  \caption{}
  \label{weighting-d}
\end{subfigure}

\begin{subfigure}{.5\columnwidth}
  \centering
  \includegraphics[width=\linewidth]{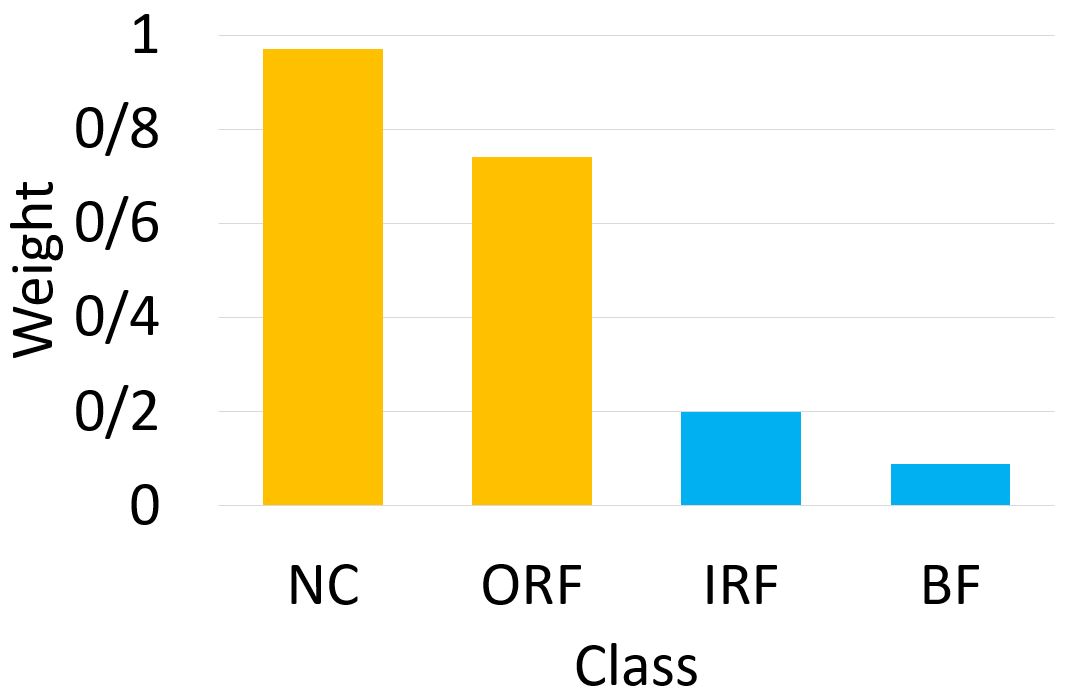}  
  \caption{}
  \label{weighting-e}
\end{subfigure}%
\begin{subfigure}{.5\columnwidth}
  \centering
  \includegraphics[width=\linewidth]{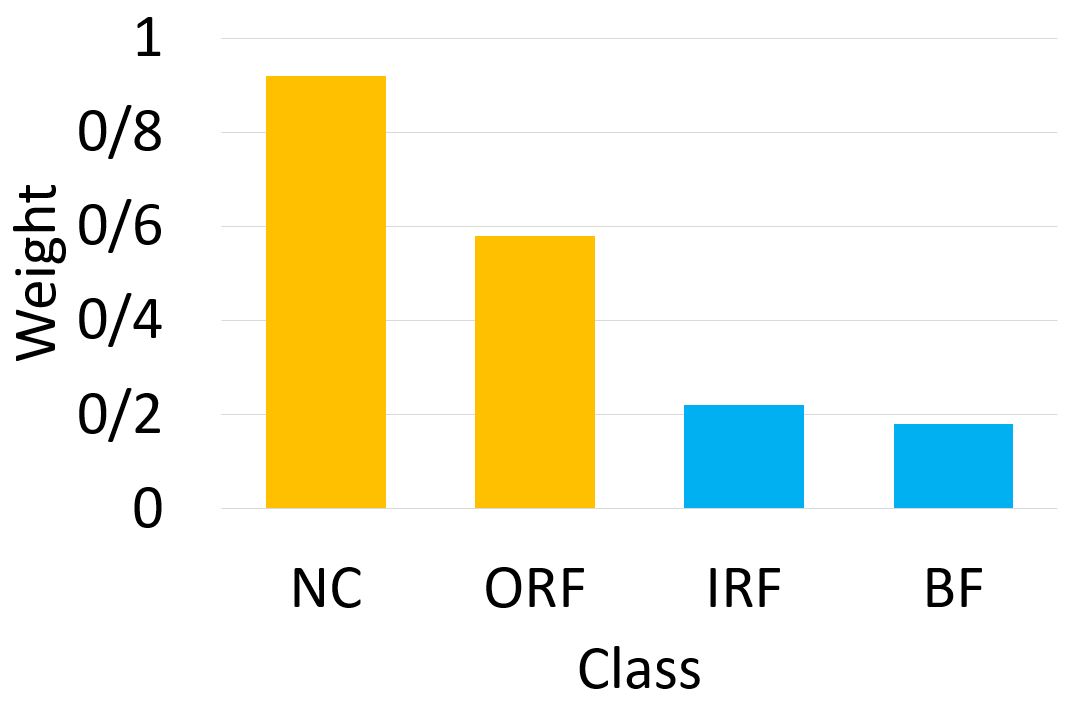}
  \caption{}
  \label{weighting-f}
\end{subfigure}
\caption{ Bar charts of each class weight . a) The PTPAI method for task C2. b) The CWATN method for task C2. c) The SDSAN method for task C2. d) The PTPAI method for task J5. e) The CWATN method for task J5. f) The SDSAN method for task J5.}
\label{weighting}
\end{figure}

The class-level weights of the PTPAI, SDSAN, and CWATN methods for tasks $C2$ and $J5$ under balanced conditions are visualized in Figure. \ref{weighting}. In the figures, yellow bars are the weights assigned to shared classes, while cyan bars are the weights allocated to outlier classes. In Figure. \ref{weighting-a}, which corresponds to task $C2$, shows that the PTPAI approach assigns weights close to 1 to shared classes, indicating accurate detection of outlier classes. On the other hand, Figure. \ref{weighting-b} shows that the CWATN method incorrectly assigns a weight of 0.4 to the outlier class ORF, which weakens its performance. Figure \ref{weighting-c} provides insights into the SDA-based method's performance in PSFD problems. Without the internal weighting mechanism, the weight assigned to the outlier classes in the source domain is substantial, leading to a significant reduction in performance for this method. Consequently, it fails to effectively mitigate the adverse effects of outlier classes. Furthermore, Figure. \ref{weighting-d} to Figure. \ref{weighting-f} illustrate the weight coefficients of different classes for the three methods in the scenario where two outlier classes are in the source domain for task $J5$ under balanced conditions. Despite the increased number of outlier classes, the PTPAI method consistently outperforms the other two methods for identifying outlier classes. This indicates the ability of the proposed approach to successfully transfer crucial information from the synthetic data of the large-scale source domain to the actual data of the small-scale target domain.

\subsubsection{Exploring the impact of different shape parameters and decision boundaries}

In this subsection, the impact of the shape parameters of the $\beta$ distribution on the RF-Mixup approach is examined. An experiment was conducted on task $J3$ in both balanced and 1\% imbalance modes to assess the effect of this hyperparameter. The outcomes of this experiment based on the F1-score criterion are depicted in Table \ref{beta}. Table \ref{beta} indicates that the proposed PTPAI approach diagnoses healthy states with high F1-score for shape parameters ranging from $\beta(0.5,0.5)$ to $\beta(2,2)$ in both balance modes. However, the performance of the PTPAI method substantially deteriorates for large shape parameters such as $\beta(200,200)$. In this case, the mixing parameter ${\lambda _e}$ tends to be around 0.5. Consequently, the RF-Mixup methodology pushes the proposed method to produce features that evenly blend the features from the domains, thereby reducing the method's generalization capability and performance specifications. Another important hyperparameter of the PTPAI method is the decision boundary of the RF-Mixup technique. To assess the impact of varying decision boundary values, an investigation was conducted for task $J5$ with balance levels of 1\% and 5\%, as presented in Table \ref{db}. Notably, the range of decision boundary values from 0.3 to 0.9 did not exhibit any significant effect on the results. However, setting the decision boundary to 0.1 resulted in a substantial decline in diagnostic performance. According to these findings, the shape parameter and decision boundary are not extremely sensitive to varied values. Nevertheless, employing an excessively large shape parameter or an exceptionally small decision boundary value can adversely affect the performance of the proposed method.


\begin{table}[th]
\caption{The effect of different shape parameters on the F1-score of task $J3$.}
\label{beta}
\resizebox{\columnwidth}{!}{%
\begin{tabular}{lcccccc}
\hline
Task          & $\beta(0.1,0.1)$ & $\beta(0.5,0.5)$ & $\beta(2,2)$ & $\beta(5,5)$ & $\beta(25,25)$ & $\beta(500,500)$ \\ \hline
J3 (1\%)      & 76.92                         & 78.76                         & 78.22                     & 78.14                     & 77.46                       & 68.40                         \\
J3 (complete) & 78.83                         & 79.24                         & 74.73                     & 68.57                     & 67.62                       & 67.14                         \\ \hline
\end{tabular}%
}
\end{table}

\begin{table}[th]
\caption{The impact of different decision boundary on the F1-score of task $J5$.}
\label{db}
\resizebox{\columnwidth}{!}{%
\begin{tabular}{lccccc}
\hline
\multirow{2}{*}{Task} & \multicolumn{5}{c}{decision boundary} \\ \cline{2-6} 
                      & 0.1   & 0.3   & 0.5   & 0.7   & 0.9   \\ \hline
J5 (1\%)              & 65.11 & 73.37 & 74.32 & 73.02 & 72.44 \\
J5 (5\%)              & 69.11 & 75.33 & 77.14 & 76.46 & 75.87 \\ \hline
\end{tabular}%
}
\end{table}

\subsubsection{Statistical test}
For a more insightful statistical examination of the outcomes achieved using the various approaches, the Friedman and post-hoc Nemenyi tests are carried out on separate tasks \cite{LEBICHOT2024123445}. To calculate the Friedman statistic, a null hypothesis states that each method is equal and has the same rank. The Nemenyi test is applicable because the null hypothesis has been rejected. The Nemenyi test indicates that the two approaches perform significantly differently if there is a critical distance (CD) between their average ranks. The CD is calculated in the following manner:
\begin{equation}
CD = c{v_\alpha }\sqrt {\frac{{\phi (\phi  + 1)}}{{6\varphi }}}
\end{equation}
where $c{v_\alpha }$ represents the critical values for post-hoc testing at the significance level $\alpha$, $\phi$ denotes the total number of methods used in the analysis, and $\varphi$ represents the number of cross-domain tasks considered in the study.
 \begin{figure}[th]

    \centering
    \includegraphics[width=\columnwidth]{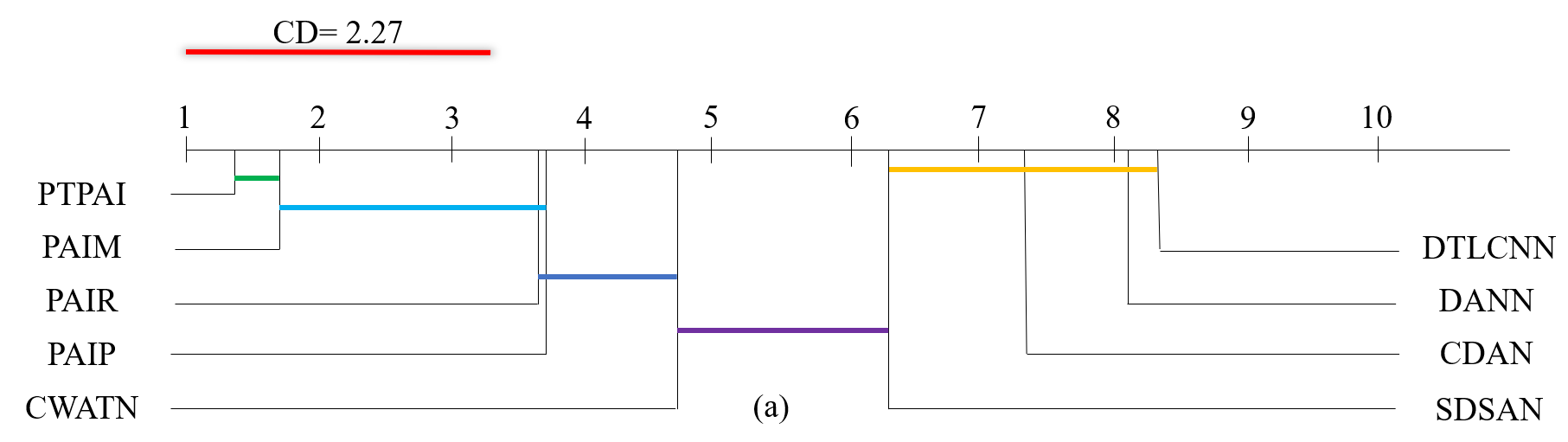}
    \includegraphics[width=\columnwidth]{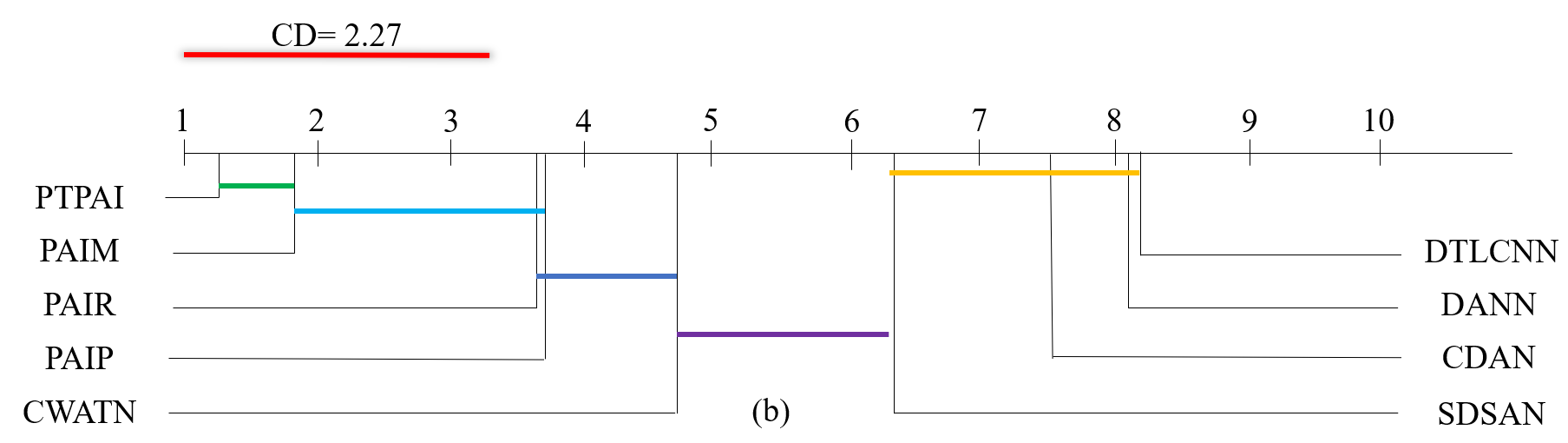}
    \includegraphics[width=\columnwidth]{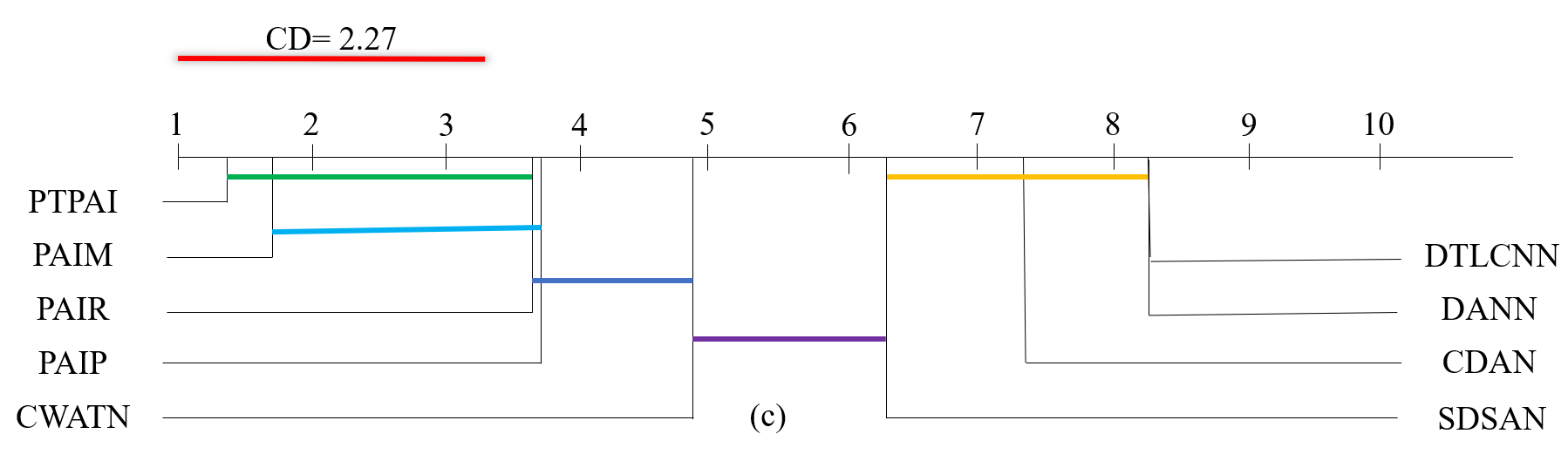}
    \includegraphics[width=\columnwidth]{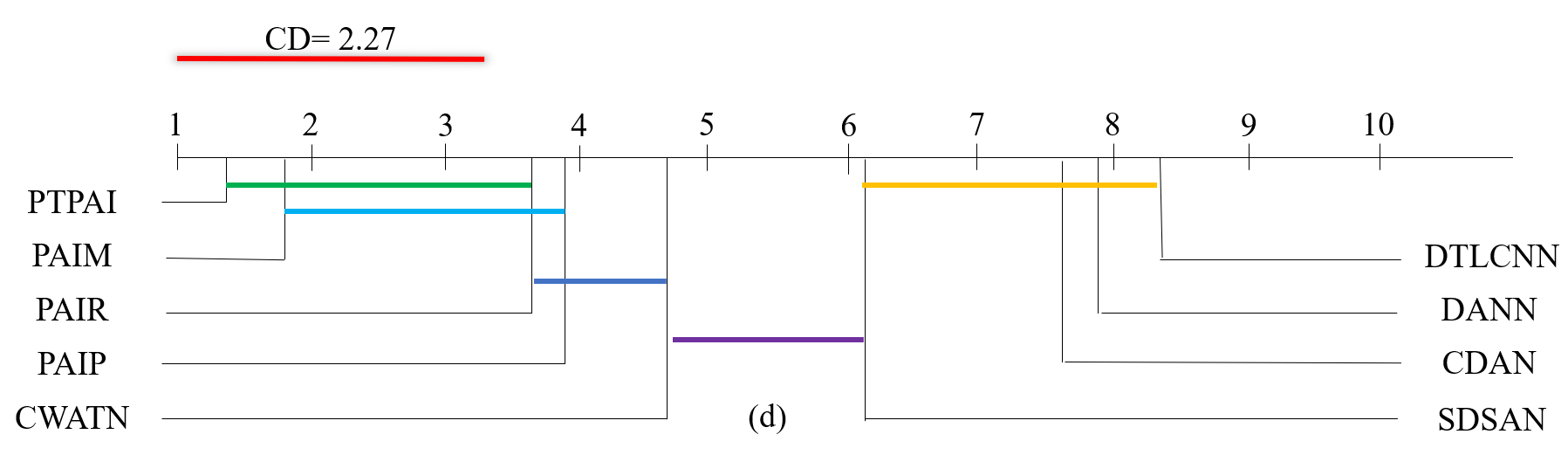}
    \caption{Critical distance representations for the Nemenyi test. The PTPAI method is compared against comparison methods in terms of b-accuracy and F1-score for CWRU and JNU datasets. (a) b-accurcy metric for CWRU, (b) b-accurcy metric for JNU, (c) F1-score metric for CWRU, (d) F1-score metric for JNU.}
    \label{CD}
    \end{figure}

\par For 9 distinct methodologies and 28 transfer tasks, the critical distance at a significance level of 0.05 is established as 2.27. Figure. \ref{CD} showcases individual CD diagrams for each dataset and evaluation metric, to connect groups of methods that do not exhibit significant differences at this level. Upon careful examination of the graphical representation, it becomes evident that the proposed PTPAI approach attains the highest ranking among the considered methodologies, surpassing its competitors with statistical significance in b-accuracy and F1-score on both the CWRU and JNU datasets. Furthermore, the HPR-based and regularization-based techniques demonstrate performance levels that are in close proximity to that of PTPAI. In contrast, the DA-based and SDA-based methods exhibit substantial disparities in performance when compared to the top four approaches across a diverse range of criteria.


\section{conclusion} \label{sec5}
This paper introduces the PTPAI method, a novel approach for bearing fault diagnosis that addresses the challenges posed by the complete absence of labeled faulty data, limited unlabeled data with significant missing values, partial-set problems, and imbalanced class distributions. To overcome the lack of labeled faulty data in real-world applications, a physics-informed method was employed to generate synthetic faulty data. This synthetic data, alongside real unlabeled data with missing values, formed the source and target domains, respectively. The presence of missing data in the target domain introduced imbalanced class and partial-set challenges, which significantly limited the efficacy of conventional domain adaptation techniques. To address these issues, the PTPAI method introduced a novel RF-Mixup regularization technique to handle imbalanced classes and incorporated weighting modules to address the partial-set problem. Experimental results on the CWRU and JNU datasets demonstrated the effectiveness of the PTPAI method in tackling these challenges, consistently outperforming existing state-of-the-art approaches. Ablation studies and analysis further emphasized the importance of the proposed method's components in addressing each specific challenge. The PTPAI method holds significant potential for achieving accurate and reliable bearing fault diagnosis in real-world applications where labeled faulty data is unavailable.

\bibliographystyle{ieeetr}
\bibliography{main.bib}

\end{document}